\documentclass[10pt,a4paper]{article}
\usepackage[utf8]{inputenc} % allow utf-8 input
\usepackage[T1]{fontenc}    % use 8-bit T1 fonts
\usepackage{hyperref}       % hyperlinks
\usepackage{url}            % simple URL typesetting
\usepackage{booktabs}       % professional-quality tables
\usepackage{amsfonts}       % blackboard math symbols
\usepackage{nicefrac}       % compact symbols for 1/2, etc.
\usepackage{microtype}      % microtypography
\usepackage{lipsum}
\usepackage{graphicx}
\usepackage{bm}
\usepackage[numbers]{natbib}
\usepackage{amsmath}
\usepackage{lmodern}
\usepackage{amsmath}
\usepackage{bm}
\usepackage{authblk}
\usepackage{tabularx}
\usepackage{makecell}
\usepackage{array}
\usepackage[margin=1in]{geometry}
\graphicspath{{figures/}} %Setting the graphicspath
\usepackage{graphicx}
\usepackage{listings,xcolor}
\usepackage{marvosym}
\usepackage{graphicx}
\usepackage[export]{adjustbox}
\usepackage{ifsym}
\newcolumntype{L}[1]{>{\raggedright\arraybackslash}p{#1}}

\newcolumntype{C}[1]{>{\centering\arraybackslash}p{#1}}

\newcolumntype{R}[1]{>{\raggedleft\arraybackslash}p{#1}}

\usepackage{float}
\usepackage{hyperref}
\DeclareMathOperator*{\argmax}{arg\,max}

\DeclareMathOperator{\vect}{vec}
\DeclareMathOperator{\sgn}{sgn}

\newcommand{\NLMEDynamicalModelFit}{\textbf{\small NLMEDynamicalModelFit}}

\title{NLMEModeling: A Wolfram Mathematica Package for Nonlinear Mixed Effects Modeling of Dynamical Systems}

\author{Jacob Leander$^{1,2,3,{\textrm{\Letter}}}$, 
Joachim Almquist$^3$, Anna Johnning$^1$, \\Julia Larsson$^{1,2}$, and Mats Jirstrand$^1$}
\date{\small
    $^1$Fraunhofer-Chalmers Centre, Chalmers Science Park, 412 88 Gothenburg, Sweden\\%
    $^2$Department of Mathematical Sciences, Chalmers University of Technology and\\ University of Gothenburg, 412 96 Gothenburg, Sweden\\
    $^3$Clinical Pharmacology \& Quantitative Pharmacology, Clinical Pharmacology \& Safety Sciences, \\ AstraZeneca R\&D, Gothenburg, Sweden\\
    [2ex]%
    \Letter { }\href{mailto:jacob.leander@fcc.chalmers.se}{jacob.leander@fcc.chalmers.se}
}

\providecommand{\keywords}[1]
{
  \small	
  \textbf{Keywords:} #1
}

\begin{document}

\maketitle

\begin{abstract}
Nonlinear mixed effects modeling is a powerful tool when analyzing data from several entities in an experiment. In this paper, we present NLMEModeling, a package for mixed effects modeling in Wolfram Mathematica. NLMEModeling supports mixed effects modeling of dynamical systems where the underlying dynamics are described by either ordinary or stochastic differential equations combined with a flexible observation error model. Moreover, NLMEModeling is a user-friendly package with functionality for model validation, visual predictive checks and simulation capabilities. The package is freely available and provides a flexible add-on to Wolfram Mathematica.
\\\\
\noindent
\end{abstract}

\section{Introduction}
In several applications, repeated measurements are collected from a number of entities to study a specific system of interest. The nonlinear mixed effects (NLME) model is a popular statistical framework able to quantify variability in response, especially popular in pharmacometrics and drug development. Typically, the underlying system of interest is described by a system of ordinary differential equations in combination with an observation model. In recent years there has been an increasing interest in extending the NLME framework to incorporate stochastic differential equations, leading to a class of models called stochastic differential equations mixed effects models (SDEMEMs) \cite{Overgaard2005,Mortensen2007,Picchini2010,Picchini2011,Delattre2013,Matzuka2016}. There are several software options available for parameter estimation in NLME models with ordinary differential equations, including both commercial software such as NONMEM \cite{Bauer2019a,Bauer2019,NONMEM74}, Monolix \cite{Monolix}, and Phoenix, and open-source such as nlmixr \cite{Fidler2019,Schoemaker2019} and Stan \cite{Carpenter2017}. However, the number of options available for SDEMEMs is limited \cite{Klim2009,Tornoe2005,Dion2019}. A programming platform well suited for modeling and scientific computation is Wolfram Mathematica \cite{Mathematica} but currently there is no functionality for performing mixed effects modeling in Wolfram Mathematica. 

In this paper, we present the software package NLMEModeling used for NLME modeling of dynamical systems in Wolfram Mathematica. The NLMEModeling package provides an easy-to-use, integrated NLME modeling environment in Wolfram Mathematica. The current version supports dynamical models with mixed effects where the dynamical system is described by either ordinary or stochastic differential equations. By utilizing the symbolic calculation capabilities and compact syntax in Wolfram Mathematica, a user-friendly package is provided. Moreover, the user can develop additional functionality to tailor the package towards its own needs. The modeling framework has previously been applied in several applications, including oncology \cite{Cardilin2017,Cardilin2018,Cardilin2019}, single cell experiments \cite{Almquist2015a}, and pharmacokinetic and pharmacodynamic modeling \cite{Tapani2014,Leander2015,Andersson2016,Andersson2017}, but the package has not previously been publicly available. The purpose of this paper is to present the NLMEModeling package from a user perspective and to introduce interested users of Wolfram Mathematica to the field of NLME modeling, but not to compare the package with existing software in terms of computational speed and robustness. 

The paper is organized in the following way. 
First, the mathematical theory behind the mixed effects framework based on ordinary and stochastic differential equations is introduced.
Second, the functionality of the NLMEModeling package is illustrated with three examples of different complexity.
The first example is a pharmacokinetic (PK) model which serves to introduce the model syntax, parameter estimation, and model validation.
The more advanced examples consist of pharmacokinetic-pharmacodynamic models with multiple observation variables and random effects, in an ordinary as well as a stochastic differential equation setting.
We also introduce the concept of so called phi-parameters, which is a convenient way of introducing the relation between different parameters types in a mixed effects model.
Finally, the modeling package is discussed and future extensions are outlined.

\section{Methods}  
The mathematical foundations of NLMEModeling will be provided by first introducing the NLME model. Second, we describe how the estimation of model parameters is done according to the maximum likelihood approach and give a high-level description of the numerical methods used. A more detailed description of the methodology can be found in \cite{Almquist2015,Olafsdottir2018}. Instructions for retrieving and installing the package are also provided.
\subsection{The Nonlinear Mixed Effects Modeling Framework}
\subsubsection*{The statistical model}
We consider models where the underlying system is described by either a system of ordinary differential equations (ODEs) or stochastic differential equations (SDEs). In the case of ODEs, the dynamical system for individual $i$ is governed by the ordinary differential equation 
\begin{align}
\label{eq:ODE}
d\bm{x}_i &= \bm{f}(\bm{x}_i,t,\bm{u}_i,\bm{\theta},\bm{\eta}_i)dt, \quad \bm{x}_i(t_0)=\bm{x}_0(\bm{u}_i,\bm{\theta},\bm{\eta}_i),
\end{align}
\noindent
where $\bm{x}_i$ is a vector of state variables, $\bm{u}_i$ is a vector of system input or covariates, $\bm{\theta}$ a vector of fixed effects parameters, and $\bm{\eta}_i$ a vector of random effects. The random effects $\bm{\eta}_i$ are assumed to be multivariate normal distributed with mean zero and covariance matrix $\bm{\Omega}$,
\begin{equation}
\bm{\eta}_i \sim N(\bm{0},\bm{\Omega}).
\end{equation}
\noindent
In the case of stochastic dynamics, the underlying model is described by an SDE on the form
\begin{align}
\label{eq:SDE}
d\bm{x}_i &= \bm{f}(\bm{x}_i,t,\bm{u}_i,\bm{\theta},\bm{\eta}_i)dt + \bm{G}(\bm{x}_i,t,\bm{u}_i,\bm{\theta},\bm{\eta}_i)d\bm{W}_i, \quad \bm{x}_i(t_0)=\bm{x}_0(\bm{u}_i,\bm{\theta},\bm{\eta}_i),
\end{align}
\noindent
where $\bm{G}$ is a weighting matrix and $\bm{W}_i$ is a standard Wiener process with increments $d\bm{W}_i \sim N(\bm{0},dt\bm{I})$ with $\bm{I}$ being the identity matrix.  A model for the observations of the dynamical system is given by 
\begin{align}
\bm{y}_{ij} &= \bm{h}(\bm{x}_i,\bm{u}_i,t_{ij},\bm{\theta},\bm{\eta}_i) + \bm{e}_{ij},
\end{align}
where the vector $\bm{y}_{ij}$ denotes the $j$th observation for the $i$th individual. We let $N$ denote the total number of individuals and $n_i$ denote the total number of observations for individual $i$. In the observation model $\bm{e}_{ij}$ are assumed to be multivariate normal distributed according to $\bm{e}_{ij} \sim N(\bm{0},\bm{\mathrm{\Sigma}}(\bm{x}_i,t_{ij},\bm{u}_i,\bm{\theta},\bm{\eta}_i))$.

\subsubsection*{Derivation of the likelihood function}
Given an NLME model and a set of observations, we are interested in estimating the fixed effects $\bm{\theta}$, the random effects covariance matrix $\bm{\Omega}$, and the observation error covariance matrix $\bm{\Sigma}$. To simplify the notation, we use $\bm{\theta}$ to denote all parameters of interest, including parameters in $\bm{\Sigma}$, $\bm{\Omega}$, and, in the case of stochastic dynamics, also the matrix $\bm{G}$.

NLMEModeling estimates the model parameters using a method where the likelihood is approximated using a second-order Taylor expansion. The approximation of the likelihood has been derived previously \cite{Overgaard2005,Leander2015,Wang2007} and we will, therefore, only give a brief overview of the derivation.

We let $\mathcal{Y}_{i,n_i} = \{\bm{v}_{i,1},\bm{v}_{i,2},\dots,\bm{v}_{i,n_i}\}$ denote the collection of all observations for individual $i$ and let $\mathcal{Y} = \{\mathcal{Y}_{1,n_1},\mathcal{Y}_{2,n_2},\dots,\mathcal{Y}_{N,n_N}\}$ denote the totality of observations. The residuals $\bm{\epsilon}_{ij}$ are defined as 
\begin{equation}
\bm{\epsilon}_{ij} = \bm{v}_{ij} - \hat{\bm{y}}_{ij}
\end{equation}
where the expected observation value $\hat{\bm{y}}_{ij}$ and covariance $\bm{R}_{ij}$ are given by
\begin{align}
\hat{\bm{y}}_{ij} &= E[\bm{y}_{ij}|\mathcal{Y}_{i(j-1)}] \\
\bm{R}_{ij} &= Cov[\bm{y}_{ij}|\mathcal{Y}_{i(j-1)}]
\end{align}
\noindent
For ODE models, $\bm{R}_{ij}$ is equal to the observation error covariance matrix $\bm{\Sigma}$ as the dynamical model is deterministic. For SDE models, on the other hand, NLMEModeling utilizes the extended Kalman filter \cite{Jazwinsky1970} to estimate $\hat{\bm{y}}_{ij}$ and $\bm{R}_{ij}$ \cite{Leander2015,Olafsdottir2018,Leander2014}. 

Since the random effects are unobserved quantities, we marginalize over the random effects to obtain an expression of the likelihood function that depends only on $\bm{\theta}$. 
Assuming independence between individuals, we have
\begin{equation}
\label{eq:exactAPL}
L(\bm{\theta}|\mathcal{Y})=\prod^{N}_{i=1} \int p(\mathcal{Y}_{i,n_i}|\bm{\theta},\bm{\eta}_i)p(\bm{\eta}_i)d\bm{\eta}_i = \prod^{N}_{i=1}\int \exp(l_i)d\bm{\eta}_i.
\end{equation}
In the expression above, $l_i = l_i(\bm{\eta}_i)$ is the \emph{a~posteriori} log-likelihood for the random effects of the $i$th individual given by
\begin{equation}
\label{Eq:iLoglh}
l_i=-\frac{1}{2} \sum_{j=1}^{n_i}\Big(\bm{\epsilon}_{ij}^T\bm{R}^{-1}_{ij}\bm{\epsilon}_{ij}+ \log\det \big(2\pi \bm{R}_{ij}\big)\Big)- \frac{1}{2}\bm{\eta}^{T}_{i}\bm{\Omega}^{-1}\bm{\eta}_i- \frac{1}{2}\log \det \big(2\pi\bm{\Omega}\big).
\end{equation}
In most cases, there is no closed-form expression for the integral in Equation \eqref{eq:exactAPL}, but it can be approximated using the Laplace approximation \cite{Wang2007,Vonesh1996}. The Laplace approximation uses a second-order Taylor expansion of $l_i$ around a point $\bm{\eta}_i^*$. Here, the point is chosen to be the value of $\bm{\eta}_i$ which maximizes the \emph{a~posteriori} log-likelihood in Equation \eqref{Eq:iLoglh},
\begin{equation}
\label{eq:etaoptimum}
\bm{\eta}_i^* = \argmax_{\bm{\eta}_i} l_i(\bm{\eta}_i).
\end{equation}
Using this $\bm{\eta}_i^*$, the approximate population likelihood, $L_L$, becomes
\begin{equation}
L(\bm{\theta}|\mathcal{Y}) \approx L_L = \prod^{N}_{i=1} \Bigg( \exp(l_i(\bm{\eta}_i^*)) \det \bigg[ \frac{-\Delta l_i(\bm{\eta}_i^*)}{2\pi} \bigg]^{-\frac{1}{2}}\Bigg),
\end{equation}
where $\Delta l_i(\bm{\eta}_i^*)$ denotes the Hessian of $l_i$ with respect to $\bm{\eta}_i$ evaluated at $\bm{\eta}_i^*$. By taking the logarithm of $L_L$, the approximate population log-likelihood becomes
\begin{equation}
\label{eq:approximatedAPL}
\log L_L = \sum^{N}_{i=1} \Bigg( l_i(\bm{\eta}_i^*)- \frac{1}{2}\log \det \bigg[ \frac{-\Delta l_i(\bm{\eta}_i^*)}{2\pi} \bigg] \Bigg).
\end{equation}
Dependent on the number of terms that is kept in the expression of the Hessian of the individual log-likelihood, the first-order conditional estimation (FOCE) and the FOCE with interaction method (FOCEI) are obtained, where the latter is used in NLMEModeling. For more information regarding the methods, we refer the reader to \cite{Leander2015,Wang2007}.

\subsection{Gradient Based Optimization}
The maximum likelihood estimate is obtained my maximizing the log-likelihood with respect to the model parameters
\begin{equation}
\label{eq:thetaoptimum}
\bm{\theta}^* = \argmax_{\bm{\theta}} \log L_L,
\end{equation}
which is achieved using a gradient-based optimization method called the Broyden-Fletcher-Goldfarb-Shanno (BGFS) algorithm \cite{Nocedal2006}. The BFGS is an iterative algorithm for solving unconstrained nonlinear optimization problems. It belongs to the family of quasi-Newton methods, which is a collection of methods that seek a stationary point of a function. It should be noted that the BGFS algorithm is a local optimization method, meaning that global optimum is not guaranteed.

In NLMEModeling the gradient of the objective function is calculated using so called exact gradients. Instead of the commonly used finite difference approach, exact gradients are calculated using the analytical solution to the gradient. The derivation requires symbolic differentiation of the model with respect to the model parameters and this is achieved using the symbolic engine in Wolfram Mathematica. For derivation of the exact gradients in NLMEModeling, we refer the interested reader to \cite{Almquist2015,Olafsdottir2018}.

To obtain the uncertainty in the estimated parameters NLMEModeling uses the variance-covariance matrix of the estimated parameters, given by the negative inverse of the Hessian matrix at the optimum.

The point $\bm{\eta}_i^*$ calculated given the optimal parameters values $\bm{\theta}^*$ is the most likely parameters for each subject. These values are referred to as Empirical Bayes Estimates (EBEs). 
\subsection{Differential Equation Solver}
To solve the system of ordinary differential equations, NLMEModeling utilizes the Mathematica NDSolve framework. NDSolve is a collection of several differential equation solvers, including adaptive solvers with the ability to automatically detect and switch solver for stiff and non-stiff problems. This is important since the solution to the ordinary differential equations might behave differently for different values of the model parameters.

For the simulation of stochastic differential equations, NLMEModeling uses the Euler-Maruyama approximation \cite{Kloeden1992}, which is implemented as part of the package.

\subsection{Model Evaluation}
Functionality for goodness-of-fit analysis and model validation is provided in NLMEModeling.
This includes population predictions versus observations, individual predictions versus observations, individual weighted residuals versus time, and individual weighted residuals versus individual predictions.
In addition, several diagnostics of the EBEs are provided as well as functionality for visual predictive checks.
The VPC functionality supports both the standard VPC plot and prediction-corrected VPC.
For additional information regarding the evaluation of nonlinear mixed effects model we refer the reader to \cite{Nguyen2017}. For details regarding the VPC methods, we refer the reader to \cite{Bergstrand2011}. 
\subsection{Installing the Package}
The latest release of NLMEModeling can be obtained from \url{http://www.fcc.chalmers.se/software/other-software/nlmemodeling} and requires an installation of Mathematica 8.0 or later. The package is installed using the installation functionality in Mathematica found under File -> Install. In the Type of Item dropdown menu, choose Application. In the Source dropdown menu, choose From File. Choose OK and the package is installed. To load the package, start a new Mathematica session and run the command
\\\\
\framebox[\textwidth]{\includegraphics[]{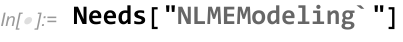}\hfill}
\section{An Introductory Example}
This section is meant to serve as a tutorial on the NLMEModeling package. The modeling workflow is illustrated using a simple PK model with the purpose of introduce the syntax, main functions, and the resulting model object. In the Advanced examples section, two additional model examples of increasing complexity are provided. 

\subsection{The Model}
In this hypothetical example we consider a one-compartmental PK model with oral administration of a drug. We use $A_1$ and $A_2$ to denote the amount of drug in the absorption and central compartment, respectively. The absorption is modelled by a first-order process with rate $k_a$ (unit $h^{-1}$), and the elimination is described by the clearance $CL$ (unit $L/h$) and volume of distribution $V$ (unit $L$). The system of ordinary differential equations is given by

\begin{align}
\frac{dA_1}{dt} &= -k_a A_1(t),\text{ }A_1(0) = Dose \\
\frac{dA_2}{dt} &= k_a A_1(t) - \frac{CL_{ind}}{V} A_2(t),\text{ }A_2(0) = 0. 
\end{align}
The individual clearance $CL_{ind}$ is log-normally distributed between individuals,

\begin{equation}
CL_{ind} = CL\exp(\eta_{1})
\end{equation}
where $\eta_{1}$ is a random effect that is normally distributed with mean 0 and variance $\omega_{1}^2$. Observations of the concentration in the central compartment,

\begin{equation}
c(t) = \frac{A_2(t)}{V},
\end{equation}
are taken at time points 0.25, 0.5, 1, 1.5, 2, 3, 4, 6, 8, 12, 18, and 24 hours after dose, and modelled according to a proportional error model

\begin{equation}
y(t) = c(t) + e_t,\text{ }e_t \sim N(0,(\sigma_{prop}c(t))^2).
\end{equation}
The population in this example consists of 20 individuals, each administrated a dose of 100 mg. Parameter values and Mathematica code for simulating the dataset are available in Appendix 1.

\subsection{Dataset Structure}
In NLMEModeling, the dataset for a population of individuals is handled by the List entity in Mathematica. To illustrate the dataset structure, we import the dataset generated in Appendix 1 and look in detail at the data for the first subject in the dataset.
\\\\
\framebox[\textwidth]{\includegraphics[]{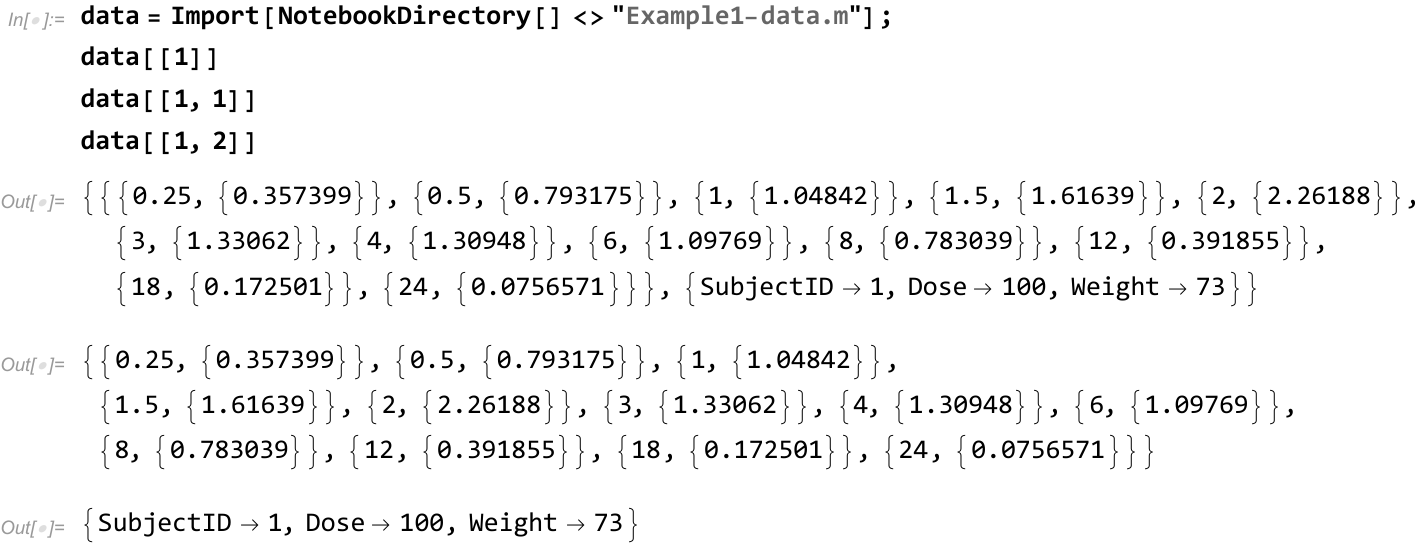}\hfill}
\\\\
As shown in the output above, the individual dataset consists of two different lists: the observation list and the rule list. Note that the observation at each time point is vector-valued, which allows for multiple variables to be measured. Moreover, the rule list is flexible and can include individual covariates (e.g. ID, weight, or sex) or more complex rules such as solutions to ordinary  differential equations (InterpolatingFunction objects) or other time-dependent expressions (e.g. dosing information).

\begin{figure}[H]
\centering
     \includegraphics[width=0.8\textwidth]{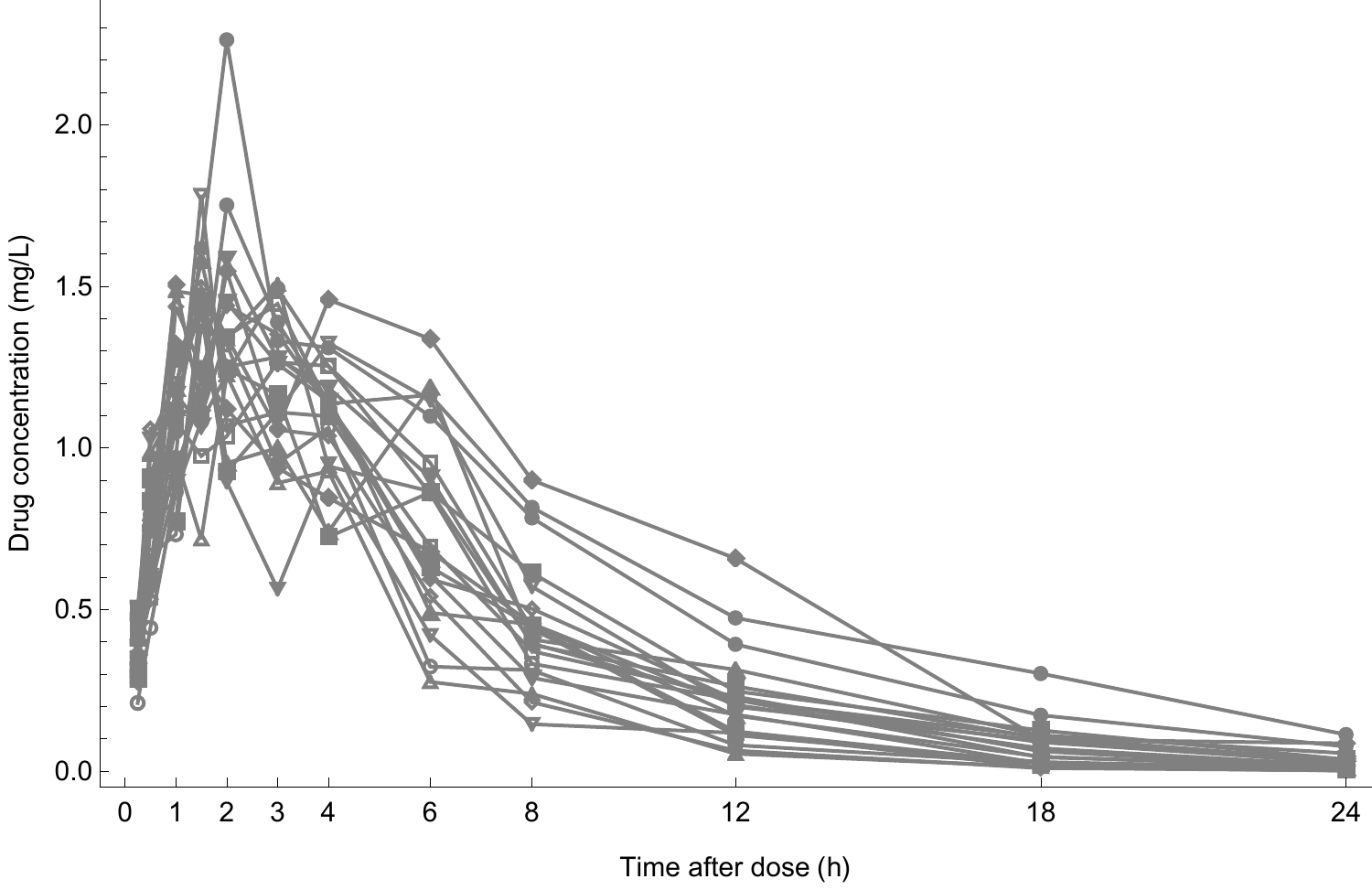}
      \caption{Simulated data for the PK experiment in Example 1.}
       \label{fig:example1-plot1}
\end{figure}

The population dataset can easily be plotted and explored using built-in functionality in NLMEModeling. It is also possible for the user of NLMEModeling to develop additional code that utilizes the pre-defined data structure. The simulated PK data for this introductory example is depicted in Figure \ref{fig:example1-plot1}.

\noindent 

\subsection{Estimating Model Parameters}
We now turn our attention to estimating model parameters given the observed data. To define the model in NLMEModeling, we need to define the dynamical model and the observation model. We will estimate model parameters using the same model structure that was used to generate the data.

We define the dynamical model as a system of ordinary differential equations using the standard notation. Note the possibility to explicitly state known relations of interest: in this example the definition of the concentration in the central compartment. We also define the observational model, here equal to the concentration in the central compartment.
\\\\
\framebox[\textwidth]{\includegraphics[]{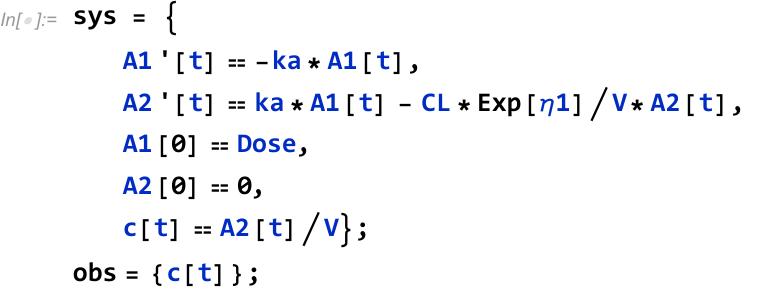}\hfill}
\\\\
The main function in the NLMEModeling package is named \NLMEDynamicalModelFit, which is used to estimate the parameters in a mixed effects model given a set of observations from a number of individuals. The minimal function call requires the data, the dynamical model, the observation model, start values for the fixed effects parameters, and a list of the random effect parameters. In this example, we are considering a proportional error model and specify this as an option (the default Sigma option is an additive error model). For a full description of the available options of \NLMEDynamicalModelFit, see Table \ref{table:optionstable}.
\\\\
\framebox[\textwidth]{\includegraphics[]{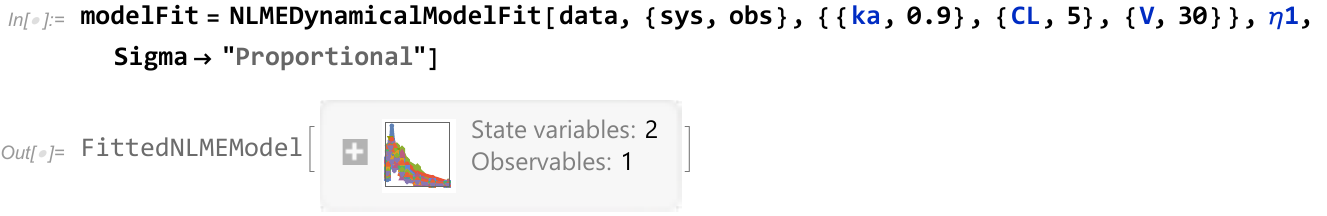}\hfill}
\\\\
Moreover, \NLMEDynamicalModelFit { is} extendable to multiple model structures within the same parameter estimation problems. In that case, the arguments is a list of datasets together with a list of system and observational models. The different models may have distinct and/or  shared random and fixed effect parameters while the random effects covariance matrix $\bm{\Omega}$ is shared between the models. For more information, we refer the reader to the package documentation.

\begin{table}[!htbp]
\caption{Options to \NLMEDynamicalModelFit. For details regarding each option, we refer the reader to the package documentation.}
\label{table:optionstable}
\centering
\begin{tabular}{|L{5cm} L{9cm}|}
\hline 
\textbf{Option name} & \textbf{Description} \\ 
\hline 
Sigma & Structure of the observation error covariance matrix $\bm{\Sigma}$. The default option is a diagonal matrix.\\ 
\hline 
Omega & Structure of the random effects covariance matrix $\bm{\Omega}$. The default option is a diagonal matrix.\\ 
\hline 
PostiveParameters & List of strictly positive parameters, for which the parameter estimation is done on log-scale to improve numerical stability.\\ 
\hline 
PerformParameterEstimation & Whether or not to perform parameter estimation\\ 
\hline 
PerformCovarianceStep & Whether or not to perform covariance calculation\\ 
\hline 
PrintToFile & Whether to print the optimization history to a file \\ 
\hline 
OuterFile & File name of the history file to be saved. \\ 
\hline 
OuterPrintOn & Specifies how often printout of optimization to the FrontEnd occurs. \\ 
\hline 
Parallel & Whether to perform the parameter estimation in parallel\\ 
\hline 
Precondition & Precondition of the covariance step. For details regarding the method, see paper by Aoki \textit{et al.} \cite{Aoki2016}  \\ 
\hline 
PrecisionGoalOuterEstimation & Number of absolute digits in the parameter estimation.\\
\hline 
RelativeDigitsOuterEstimation & Number of relative digits in parameter estimation.\\
\hline 
RelativeDigitsOuterCovariance & Number of relative digits in the covariance step. \\
\hline 
PrecisionGoalInnerEstimation & Number of absolute digits in the inner optimization.\\
\hline 
RelativeDigitsInnerEstimation & Number of relative digits in the inner optimization.\\
\hline 
PrecisionGoalInnerCovariance & Number of absolute digits in the inner optimization during the covariance step.\\
\hline 
RelativeDigitsInnerCovariance & Number of relative digits in the inner optimization during the covariance step.\\
\hline 
\end{tabular} 
\end{table}

\subsection{The Model Object}
\NLMEDynamicalModelFit{ }returns a FittedNLMEModel object, which contains information about the estimated model, including the estimated parameters, likelihood value, and model equations. See Table \ref{table:properties} for a full list of available properties that can be extracted from the model object. 

Use of the property "ModelSummary" is a convenient way to get an overview of the estimated model, including the estimated parameters values, the estimated $\bm{\Sigma}$ and $\bm{\Omega}$ matrices, and the log-likelihood value. For easy comparison between different models, $-2* \log L_L$ as well as the Akaike information criterion (AIC) and Bayesian information criterion (BIC) are provided \cite{Akaike1974,Schwarz1978}
\\\\
\framebox[\textwidth]{\includegraphics[]{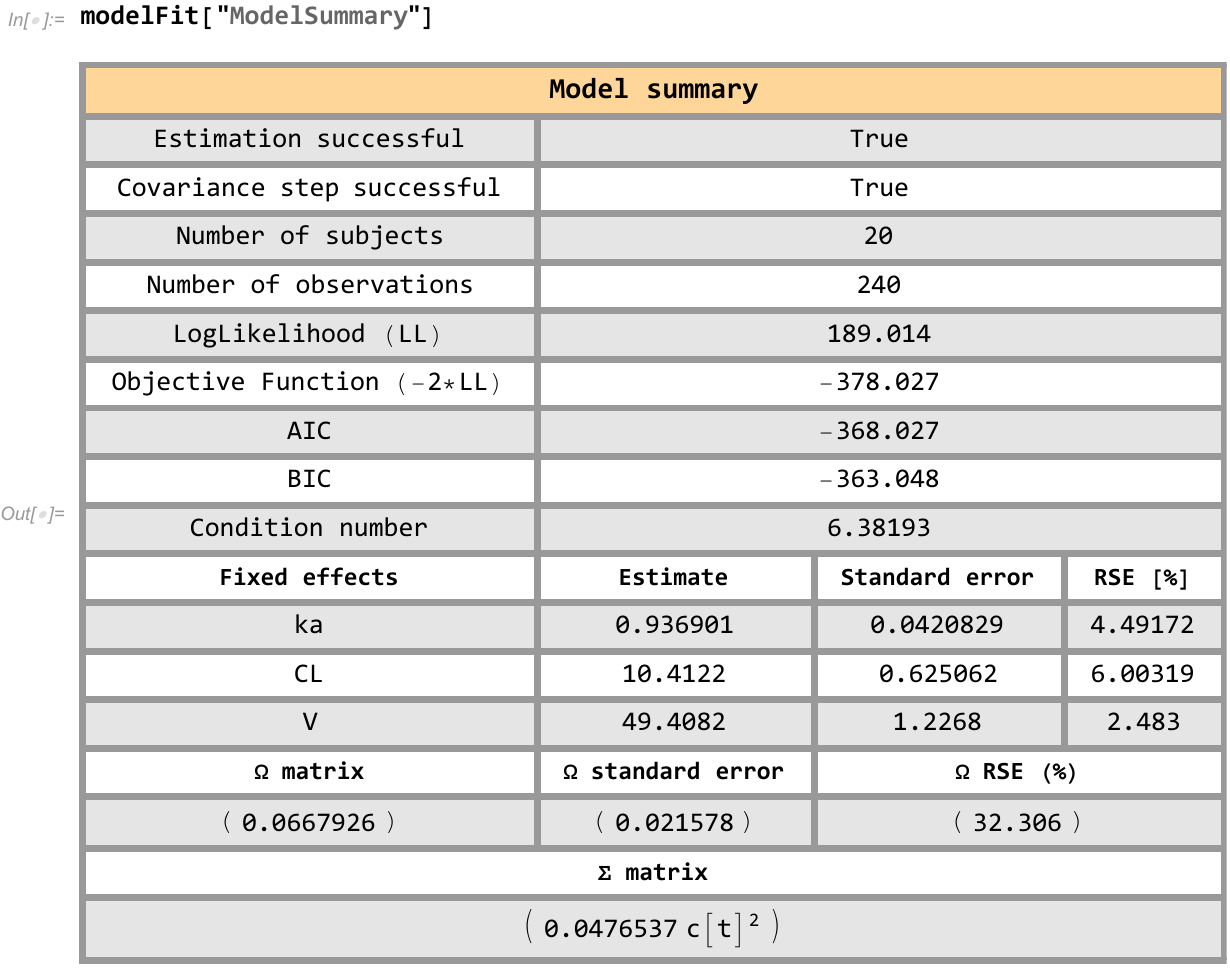}\hfill}

\begin{table}[!h]
\caption{FittedNLMEModel object properties}
\label{table:properties}
\centering
\begin{tabular}{|L{5cm}| L{9cm}|}
\hline 
\textbf{Property} & \textbf{Description} \\ 
\hline 
"ModelSummary" & A summary table of the estimated model. \\
\hline 
"NumberOfSubjects" & The total number of subjects in the datasets. \\
\hline 
"NumberOfObservations" & The total number of observations in the datasets. \\
\hline 
"ObjectiveFunctionValue" & The value of the objective function ($-2*L_L$) at the optimum. \\
\hline 
"AIC" & The Akaike information criterion. \\
\hline 
"BIC" & The Bayesian information criterion. \\
\hline 
"LogLikelihoodValue" & The value of the log-likelihood ($L_L$) at the optimum. \\
\hline 
"ConditionNumber" & The condition number of the parameter correlation matrix. \\
\hline 
"FixedEffects" & The estimated fixed effects. \\
\hline 
"OmegaMatrix" & The estimated Omega matrix. \\
\hline 
"SigmaMatrix" & The estimated Sigma matrix. \\
\hline 
"EBEs" & The empirical Bayes estimates (EBEs) for all individuals. \\
\hline 
"SuccessFlagEstimation" & True if parameter estimation succeeded, False otherwise.\\
\hline 
"SuccessFlagCovariance" & True if the covariance step succeeded, False otherwise. \\
\hline 
"EstimationTime" & The time in seconds for the estimation problem. \\
\hline 
"CovarianceTime" & The time in seconds for the covariance step. \\
\hline 
"Data" & The datasets. \\
\hline 
"Models" & The model equations.\\
\hline 
"IndependentVariable" & The independent variable in each model. \\
\hline 
"BestFit" & All estimated parameters. \\
\hline 
"ParameterCovarianceMatrix" & The covariance matrix of the parameter estimates at the optimum. \\
\hline 
"ParameterCorrelationMatrix" & The correlation matrix of the parameter estimates at the optimum.  \\
\hline 
"RandomEffectsParameters" & The symbolic random effects parameters. \\
\hline 
"PhiParameters" & The symbolic phi parameters. \\
\hline 
"StateVariables" & The state variables in each model. \\
\hline 
"WhiteNoiseVariables" & The variables corresponding to Gaussian white noise variables (empty list for ODE models). \\
\hline
"StartValues" & The start values for all parameters in the estimation problem. \\
\hline 
\end{tabular} 
\end{table}
\subsection{Model Validation}

The model object can be used by several other functions with the purpose of model validation and goodness-of-fit evaluation.

\subsubsection*{Goodness-of-fit plots}
\begin{figure}[H]
\centering
     \includegraphics[width=0.8\textwidth]{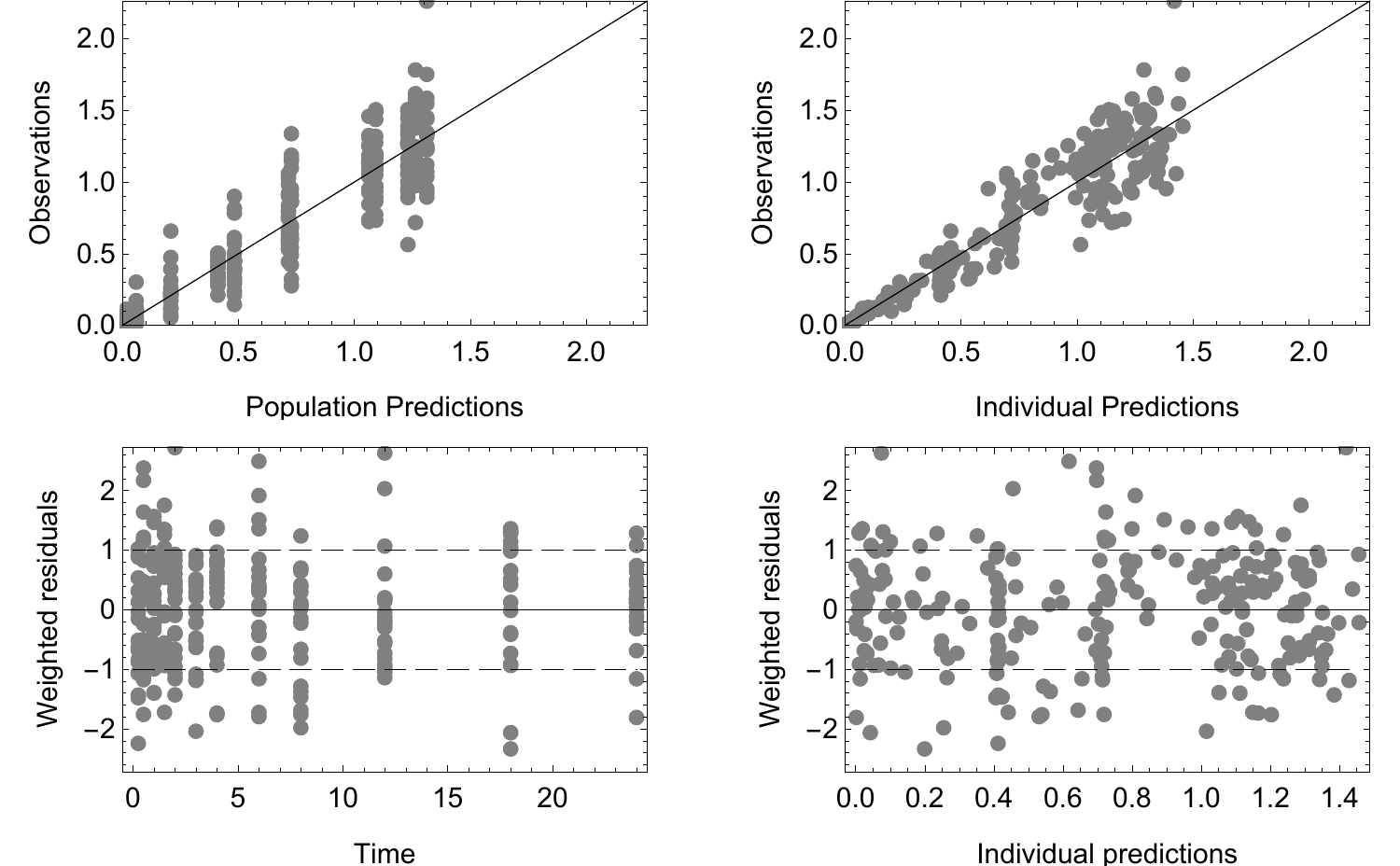}
      \caption{Basic goodness-of-fit plots for the model in Example 1.}
       \label{fig:Example1-GOFPlot}
\end{figure}

To obtain a collection of basic goodness-of-fit plots, as shown in Figure \ref{fig:Example1-GOFPlot}, we make use of the function call
\\\\
\framebox[\textwidth]{\includegraphics[]{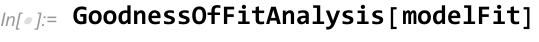}\hfill}
\\\\
The obtained figure consists of four different panels: 1) population predictions versus observations (where the random effects are set to zero), 2) individual predictions versus observations, 3) individual weighted residuals versus time, and 4) individual weighted residuals versus individual predictions.

\subsubsection*{Visual predictive checks}
\begin{figure}[H]
\centering
     \includegraphics[width=0.8\textwidth]{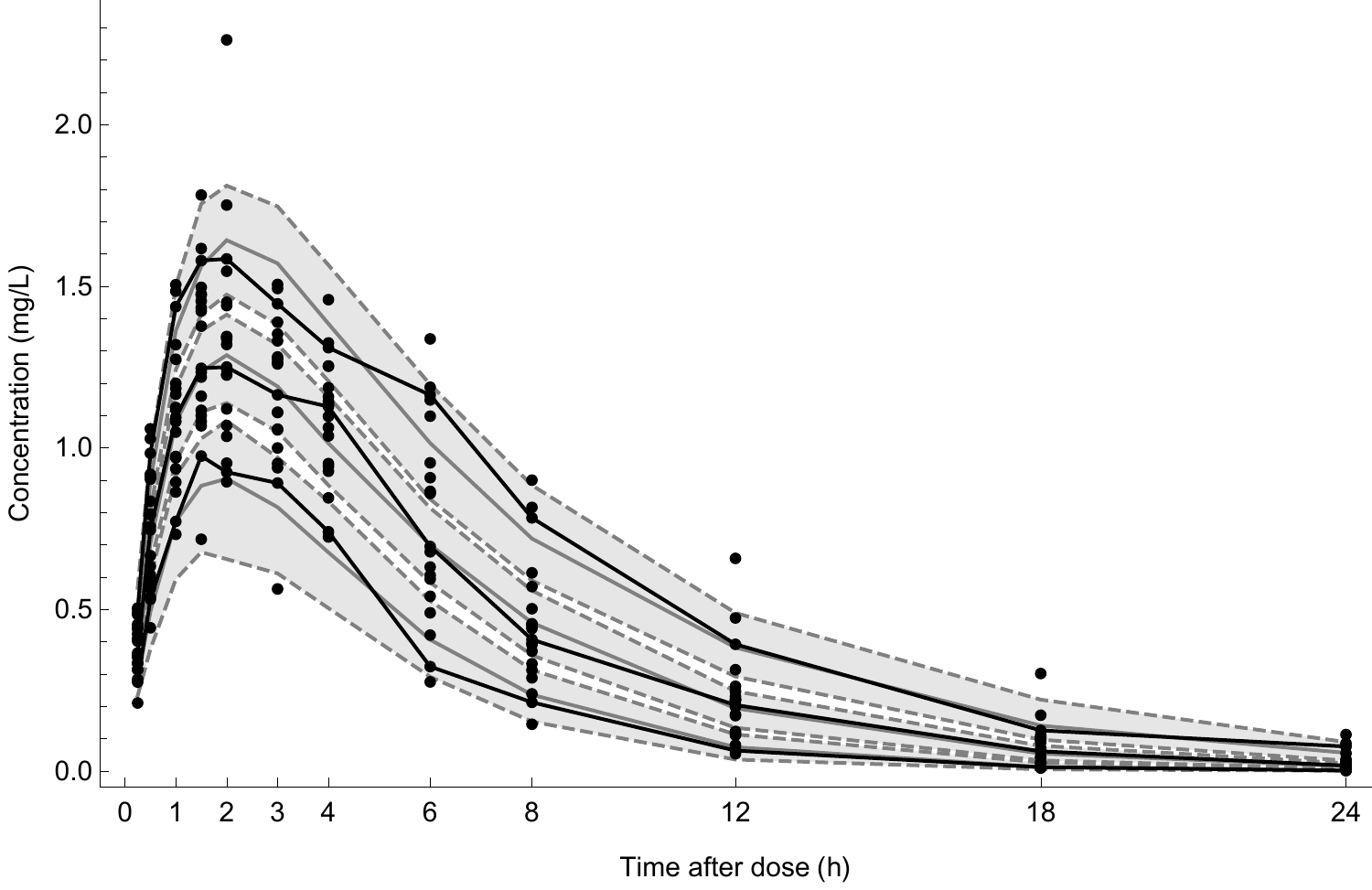}
      \caption{A VPC plot for the observations for the estimation model in Example 1.}
       \label{fig:Example1-VPCPlot}
\end{figure}

To generate a VPC based on 200 simulated datasets showing the 10th percentile, median, and 90th percentile with 90\% confidence intervals, we call
\\\\
\framebox[\textwidth]{\includegraphics[]{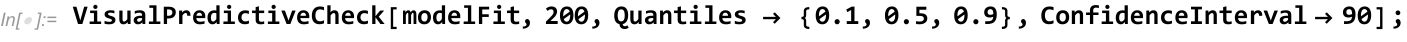}\hfill}
\\\\
\noindent
yielding the VPC plot shown in Figure \ref{fig:Example1-VPCPlot} (prediction-correction is set to False by default). The black solid lines show the calculated quantiles from observations, the grey solid lines show corresponding quantiles from the model simulation (with shaded confidence intervals), and the black dots show the observed data. As seen in VPC plot, we have good agreement between the observations and model predictions which is to be expected in this simulated example.

\section{Advanced Examples}
In this section, we extend the introductory example to allow for multiple observations as well as stochastic dynamics. We will also consider the use of so called phi-parametrization of model parameters. In contrast to the introductory example, where the combination of clearance and random effect was written in the ordinary differential equations, composite phi-parameters containing both fixed and random effects can be defined. By utilizing the composite structure of fixed and random effects, the calculation of likelihood and gradient can be further improved in terms of speed. For the exact derivation of the phi-sensitivities we refer the reader to Appendix 2. 
\subsection{Example 2: Multiple Observation Variables}
\begin{figure}[H]
\centering
     \includegraphics[width=\textwidth]{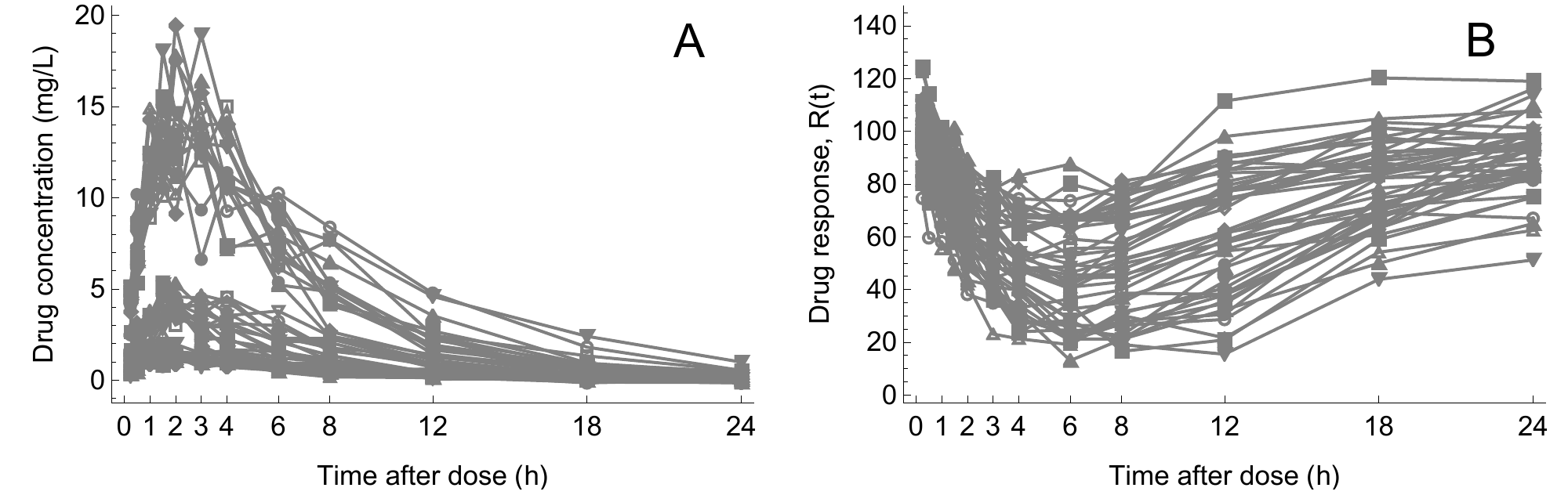}
      \caption{Simulated data from the PKPD experiment in Example 2. Panel A shows the drug concentration and panel B shows the drug response.}
       \label{fig:Example2DataPlot}
\end{figure}

Here, we give an example of a model with two observed variables. This often occurs in pharmacokinetic-pharmacodynamic (PKPD) applications where both the drug concentration and the drug effect are measured during an experiment. We use the same PK model as in the introductory example, but now with an additional variable describing a drug response dependent on the drug concentration. The system of differential equations is given by
\begin{align}
\frac{dA_1}{dt} &= -k_a A_1(t),\text{ }A_1(0) = Dose \\
\frac{dA_2}{dt} &= k_a A_1(t) - \frac{CL_{ind}}{V} A_2(t),\text{ }A_2(0) = 0 \\
\frac{dR}{dt} &= k_{out} \left( E0_{ind} \left(1-\frac{c(t)}{EC50+c(t)}\right) - R(t)\right),\text{ }R(0) = E0_{ind} .
\end{align}
In the equations above the third differential equation represents an indirect response where the drug concentration $c(t)=A_2(t)/V$ inhibits the production term. The random effects $\bm{\eta} = (\eta_{1},\eta_{2}) \sim N(0,\bm{\Omega})$ are assumed to be uncorrelated with covariance matrix

\begin{equation}
\bm{\Omega} = \begin{pmatrix}
\omega_{1}^2 & 0\\
0 & \omega_{2}^2  
\end{pmatrix}
\end{equation}
yielding log-normally distributed clearance and baseline response
\begin{align}
CL_{ind} &= CL \exp(\eta_{1}) \\
E0_{ind} &= E0 \exp(\eta_{2})
\end{align}
Compared to the introductory example we now have two types of observation variables and a slightly more complex error model: a combined error model for the PK observations and an additive error model for the PD observations. Hence, observations are assumed to be taken according to
\begin{equation}
\bm{y}(t) = (c(t),R(t)) + \bm{e}_t,\text{ } \bm{e}_t \sim N(0,\bm{\Sigma})
\end{equation}
where the observation error covariance matrix $\bm{\Sigma}$ is given by

\begin{equation}
\bm{\Sigma} = \begin{pmatrix}
\sigma_{add1}^2+(\sigma_{prop}c(t))^2 & 0\\
0 & \sigma_{add2}^2  
\end{pmatrix}.
\end{equation}
Observations of the concentration in the central compartment (unit mg/L) and of the response R(t) are taken at 0.25, 0.5, 1, 1.5, 2, 3, 4, 6, 8, 12, 18, and 24 hours after dose. In this example, we consider multiple doses (100 mg, 300 mg, and 1000 mg) with 15 subjects in each dose group. Mathematica code for simulating the dataset is available in Appendix 1. A plot of the simulated data is shown in Figure \ref{fig:Example2DataPlot}.

\begin{figure}[H]
\centering
     \includegraphics[width=0.8\textwidth]{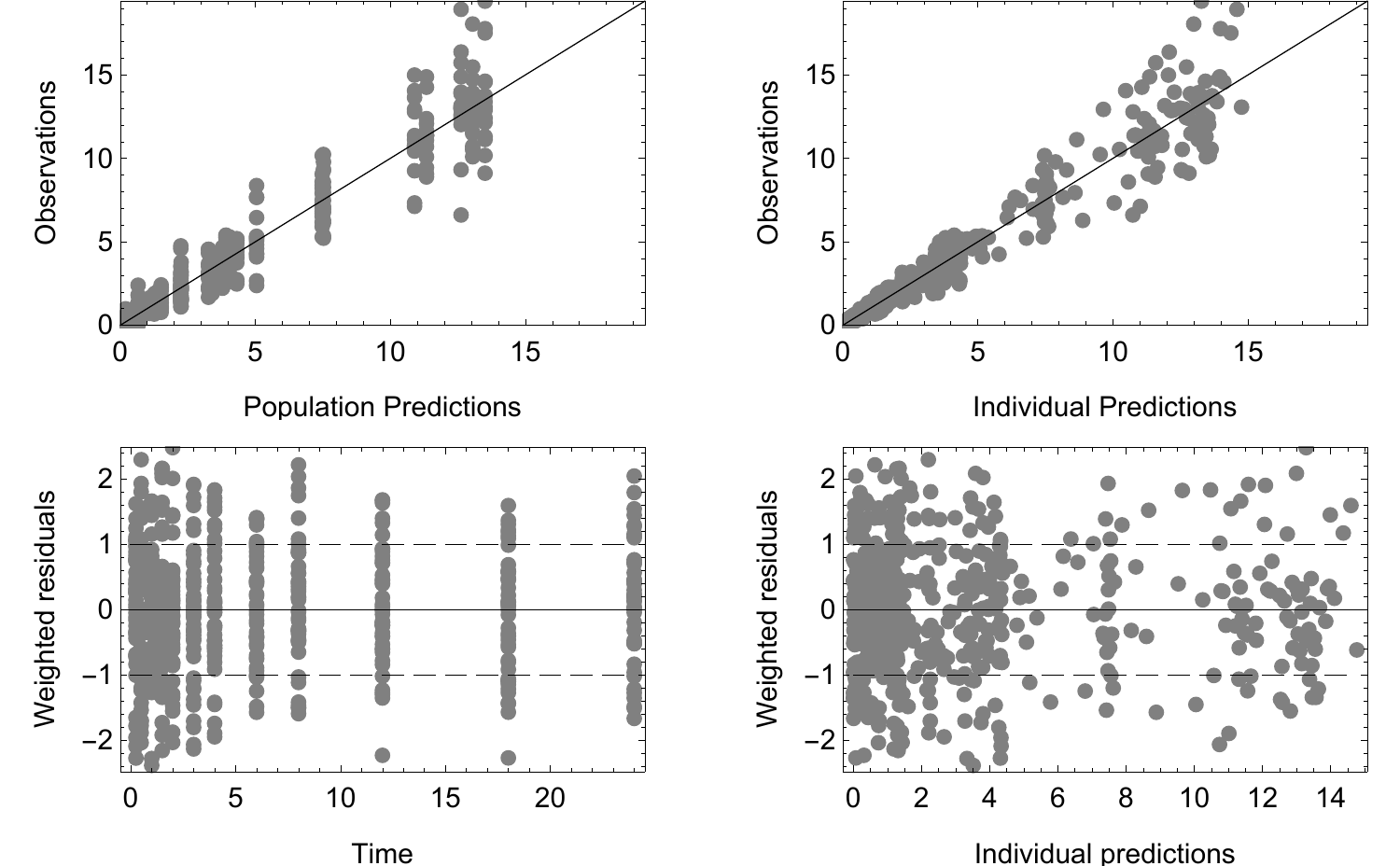}
      \caption{Standard goodness-of-fit plots for the PK observations in Example 2.}
       \label{fig:AdvancedExample1-BasicGOF-PK}
\end{figure}

\begin{figure}[H]
\centering
     \includegraphics[width=0.8\textwidth]{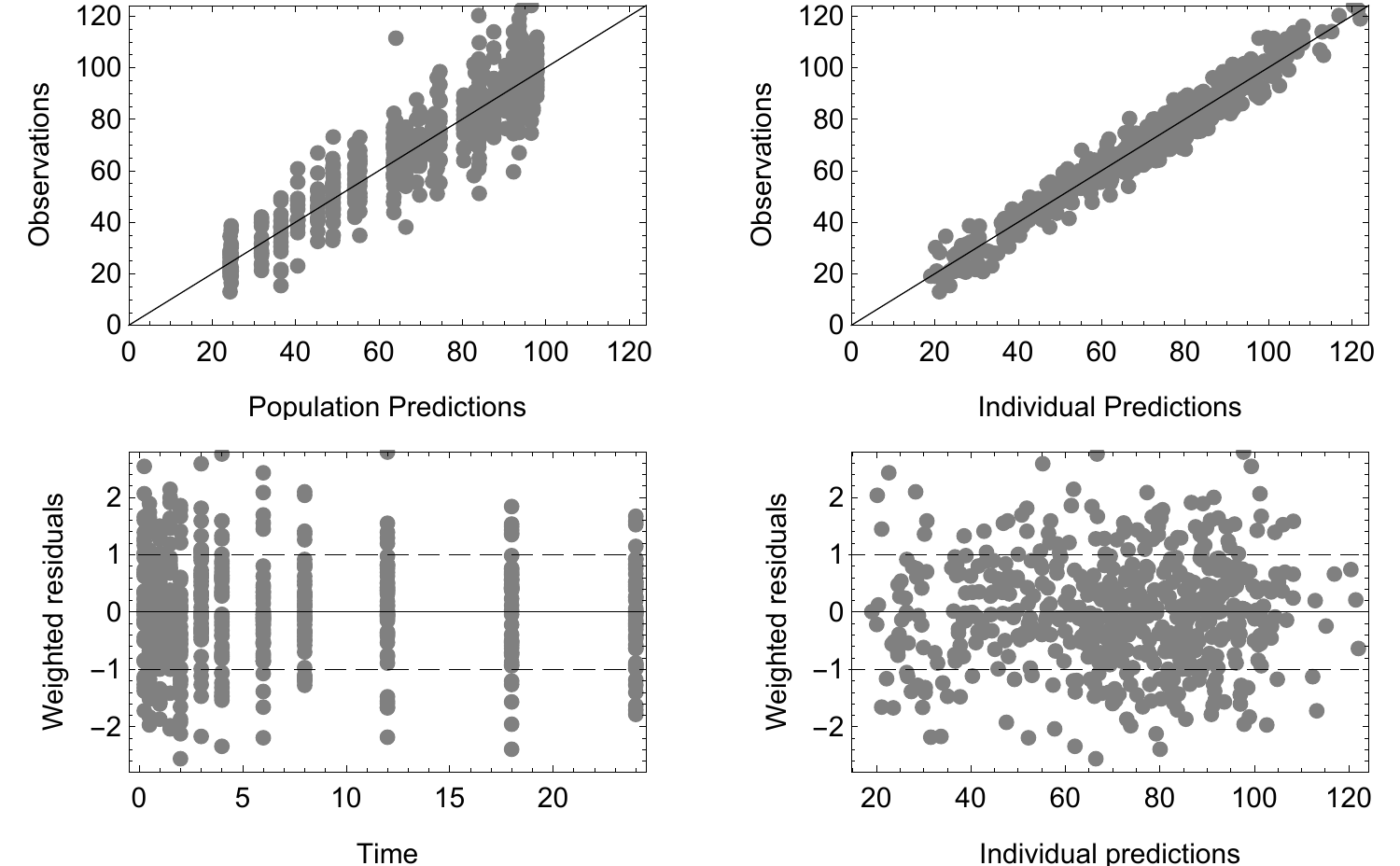}
      \caption{Standard goodness-of-fit plots for the PD observations in Example 2.}
       \label{fig:AdvancedExample1-BasicGOF-PD}
\end{figure}

\begin{figure}[H]
\centering
     \includegraphics[width=\textwidth]{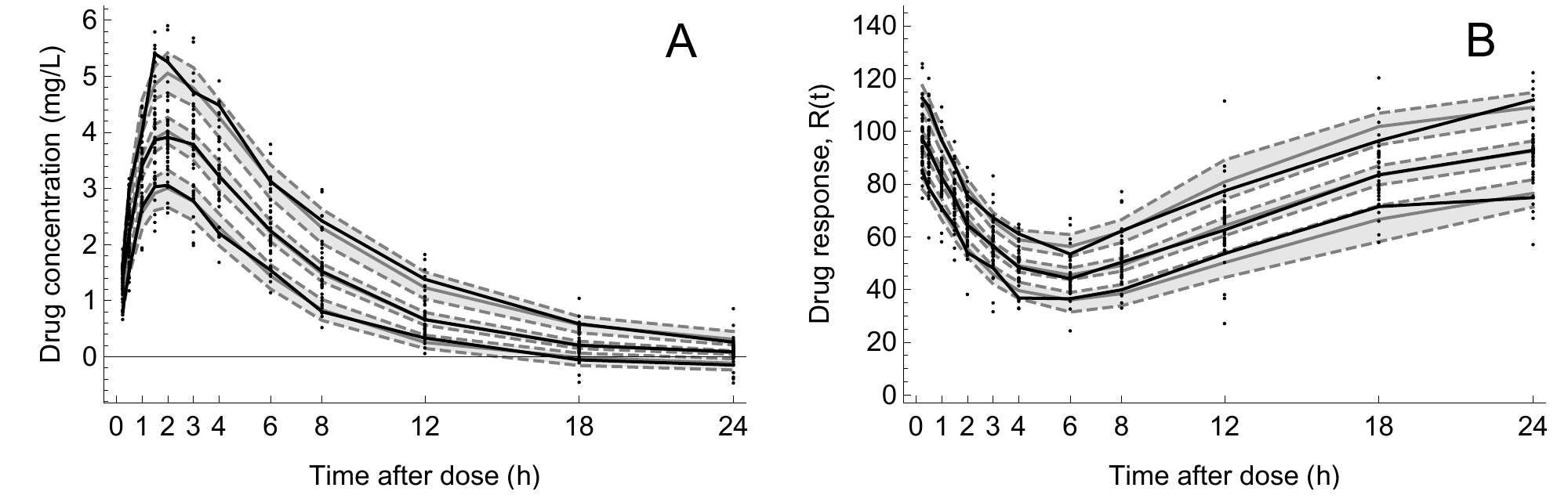}
      \caption{Prediction-corrected visual predictive checks for the PKPD model in Example 2, stratified on observation variable. Panel A shows the PK observations and panel B shows the PD observations.}
       \label{fig:Example2-VPC}
\end{figure}

\noindent
To estimate the model parameters we express the system model using an additional differential equation for the turnover model compared to the previous example. Moreover, we use direct coding of an auxiliary expression for the concentration relationship and use phi-parametrization for the inter-individual variability in clearance and baseline response. Since we measure both the concentration of the drug and the drug response, the observation is a vector consisting of two entities.
\\\\
\framebox[\textwidth]{\includegraphics[]{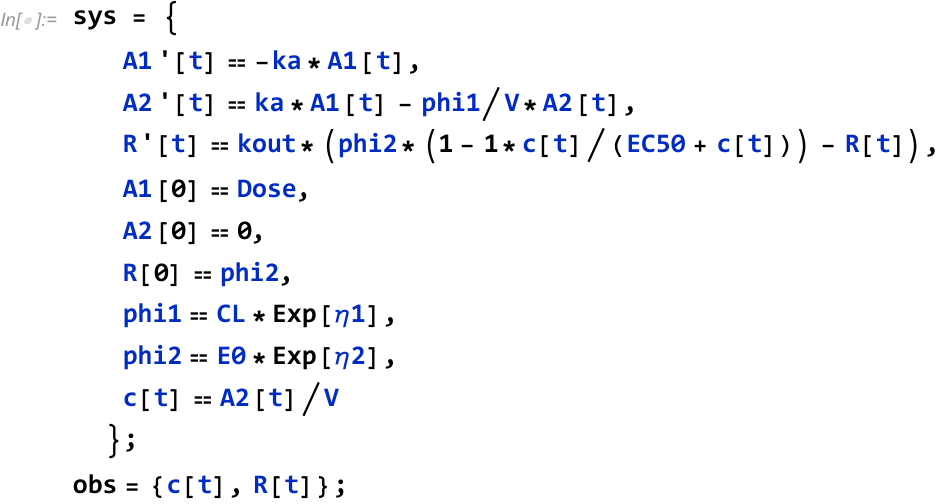}\hfill}
\\\\
In addition to the system definition, the user may also provide a user-defined structure of the random effects covariance matrix, $\bm{\Omega}$, and the observation error covariance matrix, $\bm{\Sigma}$. In this example, we consider a $\bm{\Sigma}$ matrix that describes a combined error model for the PK observations and an additive error model for the PD observations, parametrized by three parameters as follows.
\\\\
\framebox[\textwidth]{\includegraphics[]{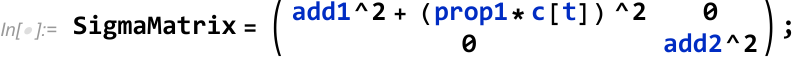}\hfill}
\\\\
To estimate the model parameters, we need the model definition, a list of the fixed effects parameters ($k_a$, $CL$, $V$, $k_{out}$, $E0$, and $EC50$) together with their corresponding initial values, and a list of the random effects parameters ($\eta_1$ and $\eta_2$). Additionally, in this example, we make use of the advanced option for the $\bm{\Sigma}$ matrix and provide the user-defined symbolic $\bm{\Sigma}$ matrix together with the related initial values. For the covariance matrix of the random effects, $\bm{\Omega}$, we here choose to estimate the full covariance matrix, specified by the option value "Full". 
\\\\
\framebox[\textwidth]{\includegraphics[]{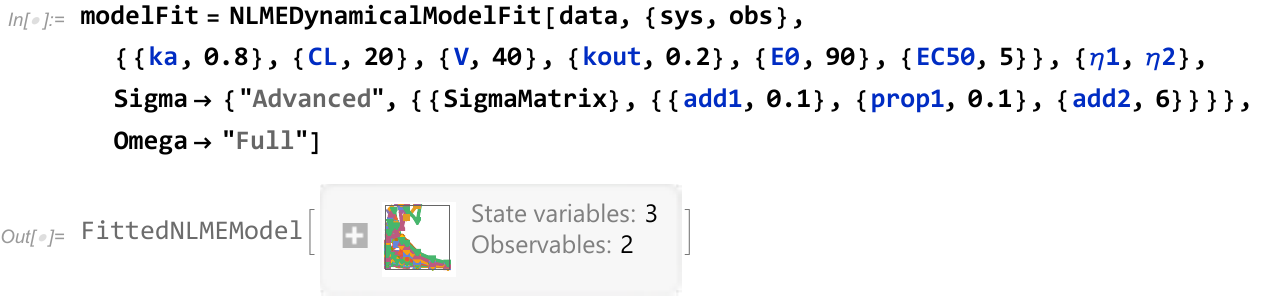}\hfill}
\\\\
The summary for the estimated PKPD model is obtained using the "ModelSummary" property.
\\\\
\framebox[\textwidth]{\includegraphics[]{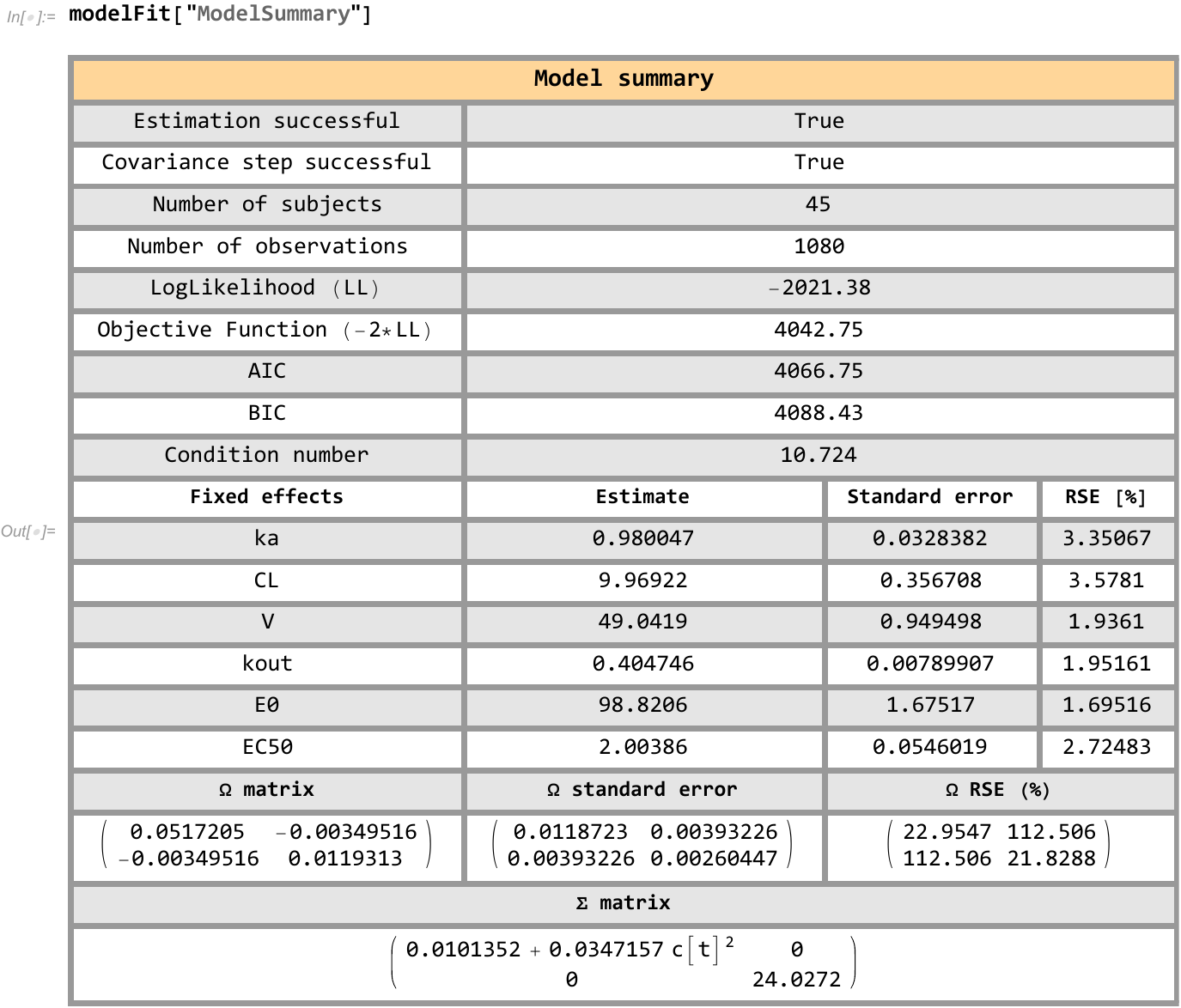}\hfill}
\\\\
The goodness-of-fit functionality extends seamlessly to multiple observation variables. When calling the GoodnessOfFitAnalysis function, a plot is created for each observation variable. The GOF plots for the PK and PD observations are shown in Figure \ref{fig:AdvancedExample1-BasicGOF-PK} and \ref{fig:AdvancedExample1-BasicGOF-PD}.
\noindent
In addition to the goodness-of-fit plots, we now turn the attention to assessment of the model assumption regarding the distribution of the random effects parameters. To validate the assumption of normality, we can plot the distribution and correlation of the Empirical Bayes Estimates (EBEs). In NLMEModeling this is achieved using the function call
\\\\
\framebox[\textwidth]{\includegraphics[]{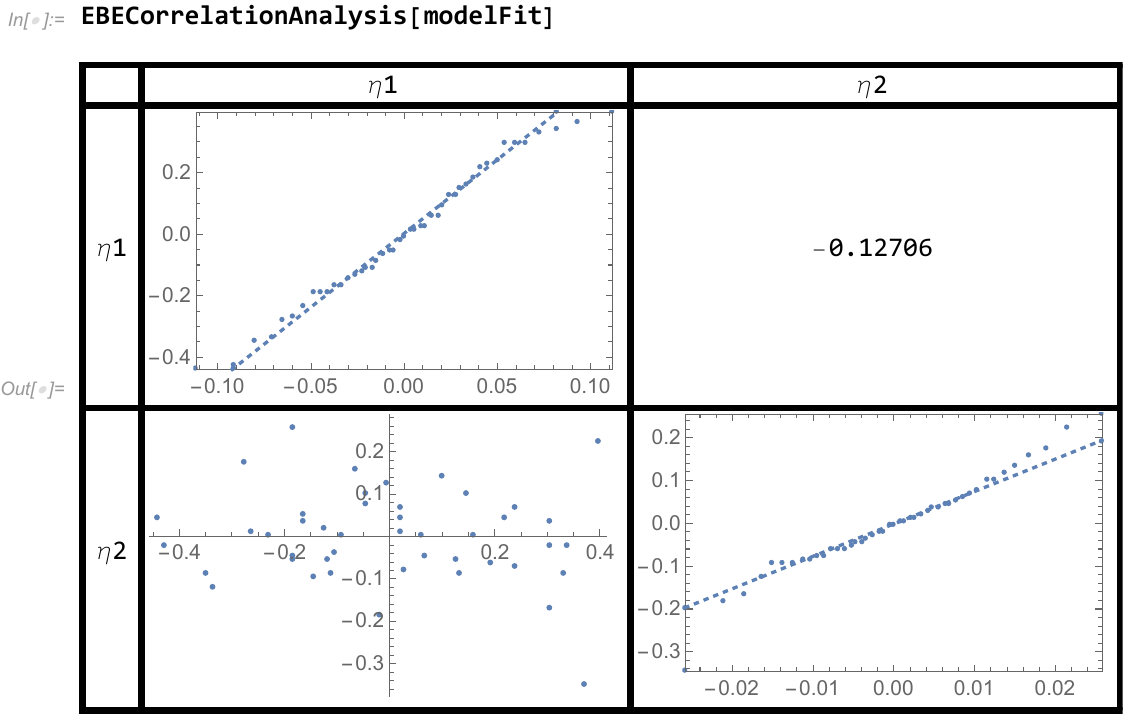}\hfill}
\\\\
where the diagonal shows the quantile-quantile plot for each random effect, the sub-diagonal plots show the pairs-plot, and the above-diagonal plots show the Pearson correlation coefficient between the EBEs. To check the distribution of the EBEs versus a specific discrete covariate of interest, we use the function EBEBoxWhiskerChart which takes the model object as argument and an option for potential stratification. The function will also report the shrinkage of each random effect parameter.
\\\\
\framebox[\textwidth]{\includegraphics[]{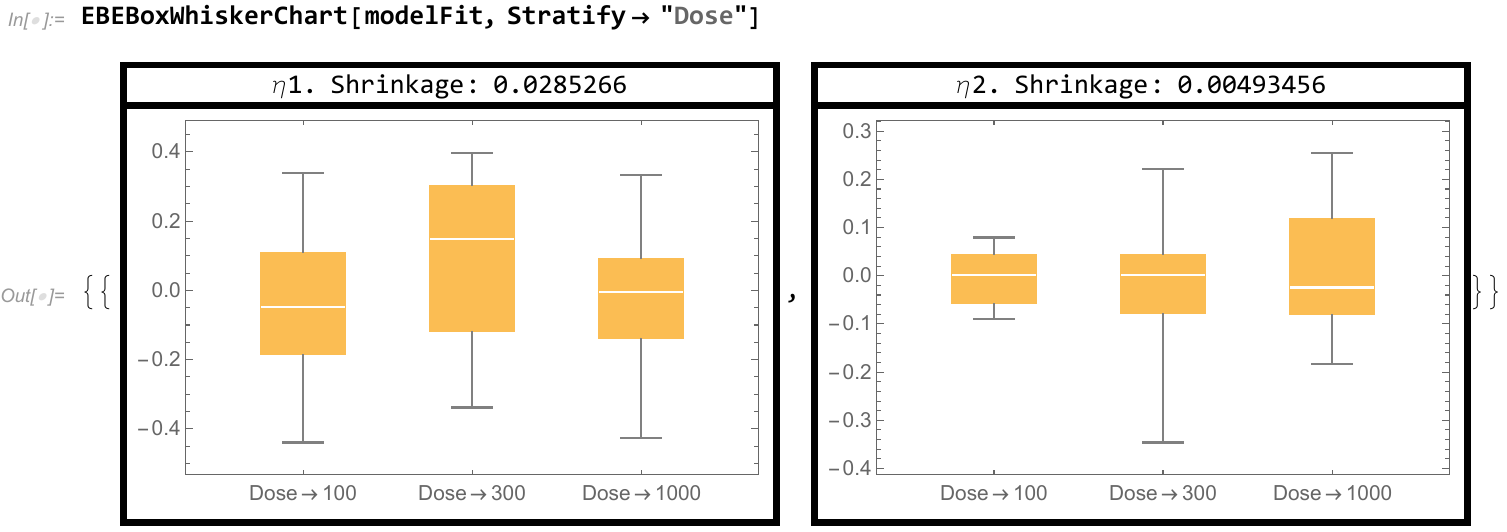}\hfill}
\\\\
The VPC analysis extends seamlessly to multiple observation variables, with a VPC for each observation variable. Here, we create a prediction-corrected VPC plot for both the PK and PD observations (using 10th, 50th and 90th percentiles with 90\% confidence interval) based on 200 simulated datasets. The resulting VPC plots are shown in Figure \ref{fig:Example2-VPC}.
\\\\
\framebox[\textwidth]{\includegraphics[]{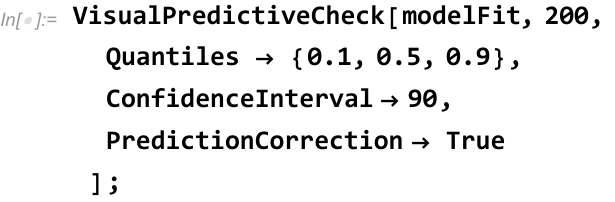}\hfill}
\\\\

\subsection{Example 3: Extension to Stochastic Dynamics}
\begin{figure}[H]
\centering
     \includegraphics[width=\textwidth]{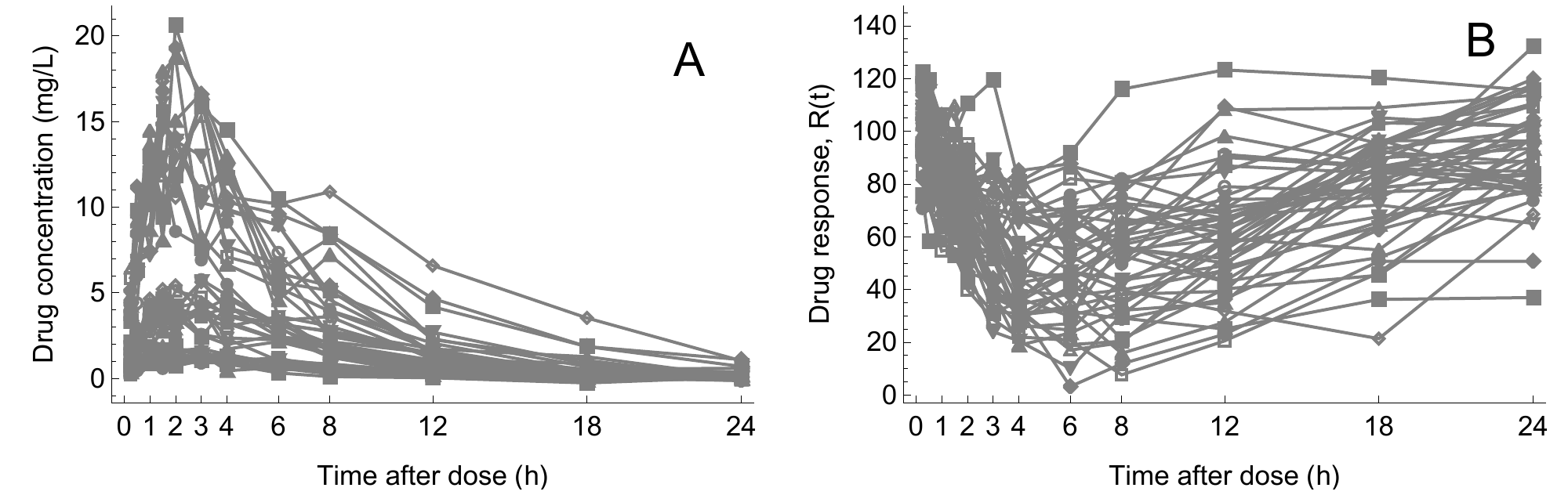}
      \caption{Simulated data from a PKPD experiment with a stochastic PD model in Example 3. Panel A shows the drug concentration and panel B shows the drug response.}
       \label{fig:AdvancedExample2-DataPlot}
\end{figure}

In this example, we illustrate how the deterministic ODE model can easily be extended to incorporate stochastic differential equations. We consider the same model as in Example 2, but replace the differential equation for the drug effect $R(t)$ with a stochastic differential equation,

\begin{align}
\frac{dA_1}{dt} &= -k_a A_1(t),\text{ }A_1(0) = Dose \\
\frac{dA_2}{dt} &= k_a A_1(t) - \frac{CL_{ind}}{V} A_2(t),\text{ }A_2(0) = 0 \\
dR &= k_{out}( E0_{ind} (1-\frac{c(t)}{EC50+c(t)}) - R(t))dt + g \cdot dW(t),\text{ }R(0) = E0_{ind}
\end{align}
where the SDE is written on differential form with $dW(t)$ being the increment of a standard Wiener process. The parameter $g$ is of special interest in this example since it quantifies the influence of the stochastic term on the dynamics. The random effects covariance matrix $\bm{\Omega}$, the observation error covariance matrix $\bm{\Sigma}$, and experimental design is kept the same as in Example 2. Mathematica code for simulating the dataset is available in Appendix 1. A plot of the simulated data is shown in Figure \ref{fig:AdvancedExample2-DataPlot}. Note the increased variability in the PD observations compared to Example 2, due to the stochastic nature of the underlying model.

\begin{figure}[H]
\centering
     \includegraphics[width=\textwidth]{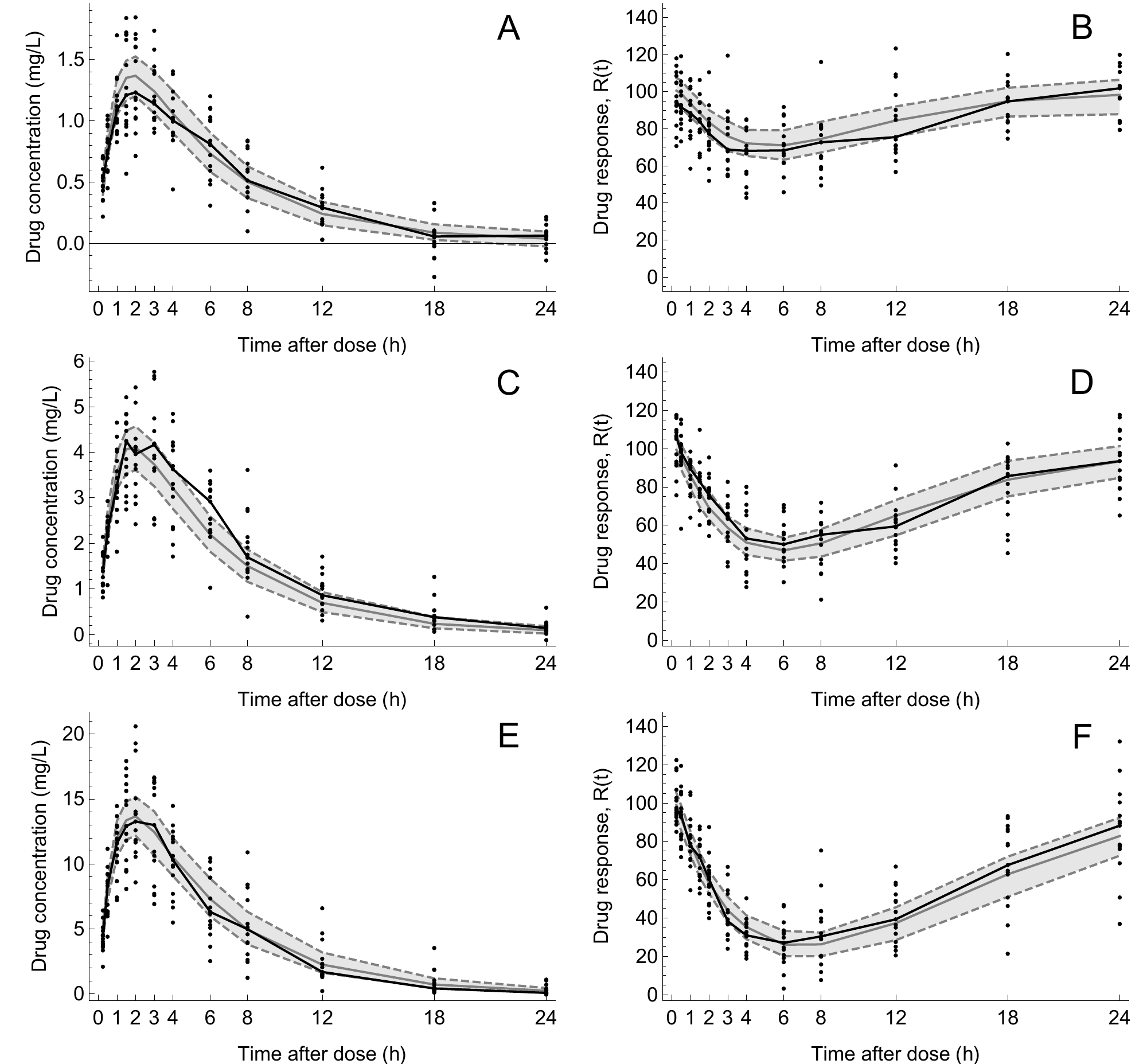}
      \caption{Visual predictive checks (VPCs) for the stochastic PKPD model in Example 3. Panel A: Dose 300 mg, PK observations. Panel B: Dose 300 mg, PD observations. Panel C: Dose 600 mg, PK observations. Panel D: Dose 600 mg, PD observations. Panel E: Dose 1000 mg, PK observations. Panel F: Dose 1000 mg, PD observations.}
       \label{fig:AdvancedExample2-VPC}
\end{figure}

To define the SDE model in NLMEModeling, the only thing we need to add is the stochastic component to the model definition. To align with the ODE notation, we here write the stochastic component as the generalized derivative of the standard Wiener process, which we will refer to as Gaussian white noise. The user can use any symbol for the Gaussian white noise (here denoted $w(t)$) and it can scale with any parameter, here multiplied with the parameter $g$.
\\\\
\framebox[\textwidth]{\includegraphics[]{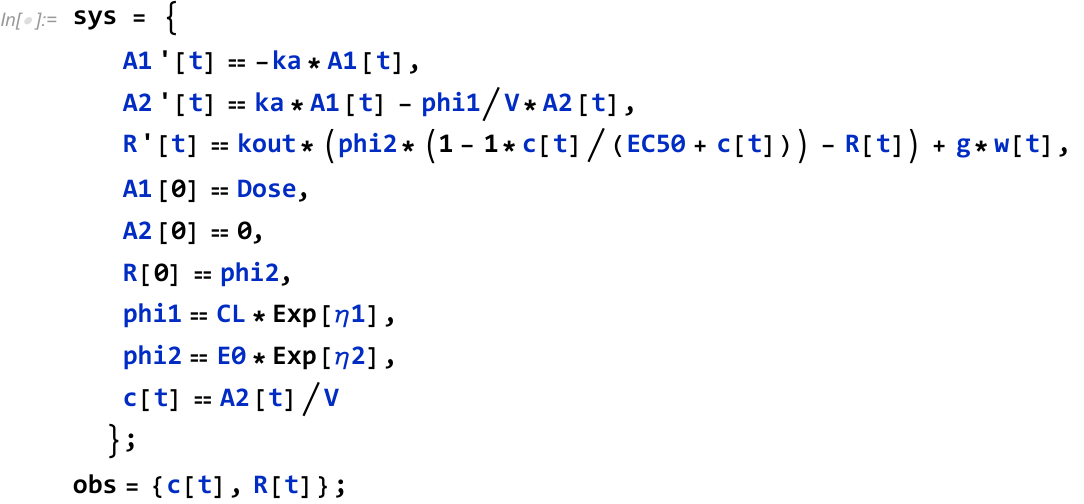}\hfill}
\\\\
To estimate the model parameters we make a similar call as for the ODE case but with an additional start value for the parameter $g$, as well as a list of the symbols used to represent the white noise variables. In this example, we use a diagonal random effects covariance matrix $\bm{\Omega}$.
\\\\
\framebox[\textwidth]{\includegraphics[]{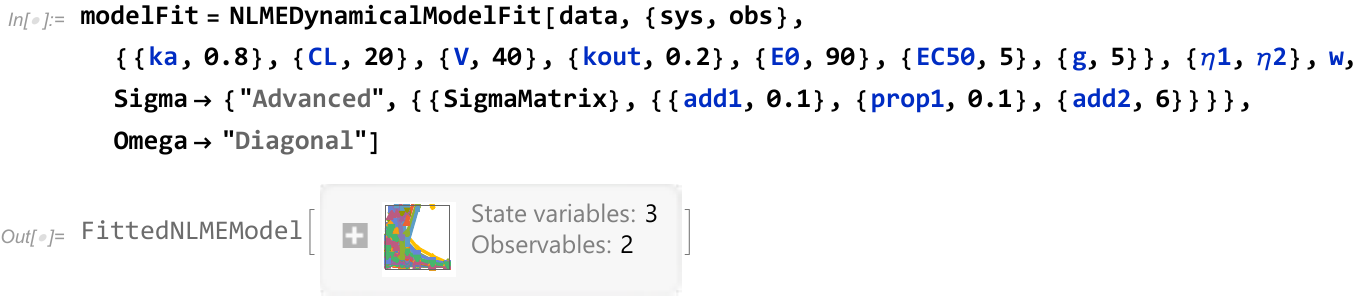}\hfill}
\\\\
The model summary for the estimated model is obtained as before, using the "ModelSummary" property. Notice the additional fixed effect parameter $g$ in the table, which is estimated with good precision (the value used for simulation was $g=10$, see Appendix 1).
\\\\
\framebox[\textwidth]{\includegraphics[]{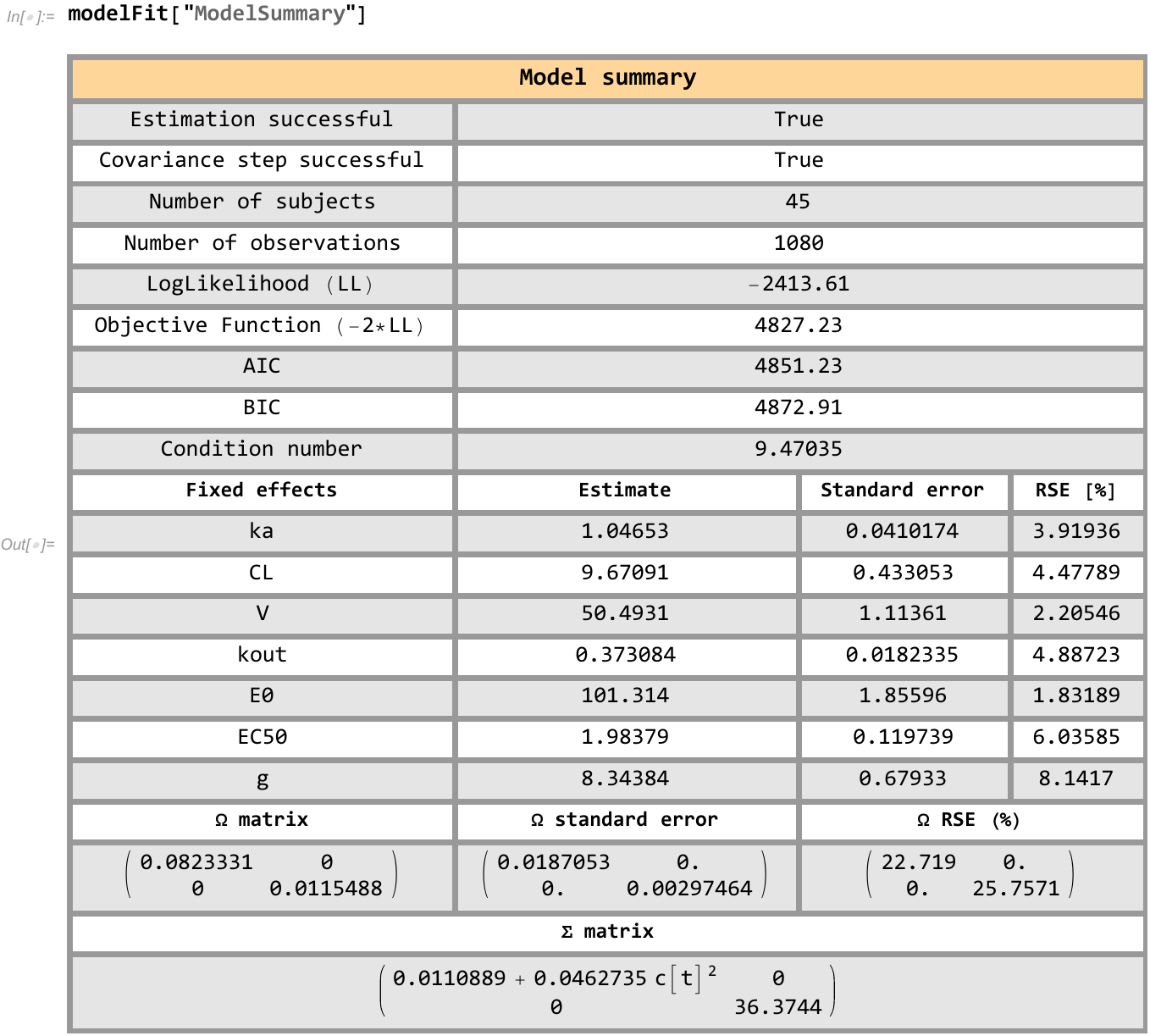}\hfill}
\\\\
The model object and functionality for stochastic models are designed to be the same as for the ODE models. As an example, we here show a visual predictive check for the stochastic mixed effects model. For each dose group, we create a VPC using the 50th percentile with 90\% confidence intervals (no prediction-correction). The resulting VPCs are shown in Figure \ref{fig:AdvancedExample2-VPC}.
\\\\
\framebox[\textwidth]{\includegraphics[]{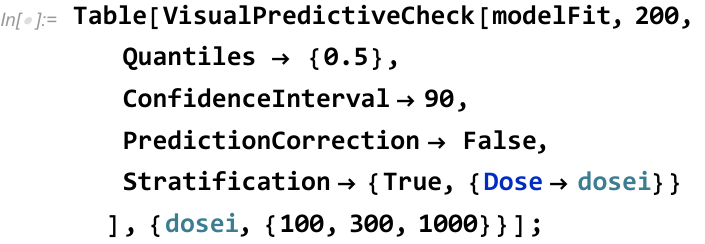}\hfill}
\\\\

\section{Discussion}
In this paper, we have presented the Wolfram Mathematica package NLMEModeling for performing nonlinear mixed effects modeling of dynamical systems. NLMEModeling offers a user-friendly environment for both simulation and estimation of nonlinear mixed effects models, and the package also comes with functionality for performing visual predictive checks and standard goodness-of-fit plots. Furthermore, the model object returned by the estimation procedure in NLMEModeling contains all the necessary information for doing additional analysis. 

NLMEModeling provides an easy-to-use modeling framework integrated into Wolfram Mathematica as an add-on package. Wolfram Mathematica  has a large library of statistical models, simulation, and visualization capabilities. By providing a mixed-effects modeling framework that is tightly linked to the Wolfram Language, the user is free to integrate other types of analyses around the mixed-effects framework presented in this paper.

In addition to the commonly used ordinary differential equations, NLMEModeling also supports stochastic differential equations mixed effects models (SDEMEMs). As discussed in the recent review paper by Irurzun-Arana \emph{et al.} \cite{IrurzunArana2020}, incorporating stochastic behavior in mixed-effects models is a promising way of capturing three sources of variability: inter-individual, intra-individual, and system stochasticity. Moreover, as the syntax and functionality for ODE and SDE models are closely linked in the package, the extension to stochastic models is seamless. To the authors best knowledge this is also the first work that provides functionality for visual predictive checks for SDEMEMs. In the current version of NLMEModeling predicted-corrected VPCs are not supported for SDEMEMs, but it is an interesting extension to the approach presented in this work. 

In terms of parameter estimation methods, NLMEModeling use the first-order conditional estimation method for approximation of the likelihood. In the case of stochastic models, the extended Kalman filter (EKF) is used for estimating the underlying state of the system conditional on the observations. The EKF has previously successfully been combined with both the FOCE method \cite{Overgaard2005,Tornoe2005} and the stochastic approximation expectation-maximization method \cite{Delattre2013}. In addition to earlier work, NLMEModeling utilizes the exact gradient method where the gradient of the objective function is calculated using the sensitivity equations of the underlying system, leading to a faster and more precise optimization routine, \cite{Almquist2015,Olafsdottir2018}.

In terms of limitations, the current release of NLMEModeling estimates the model parameters using an approximate likelihood approach. It might be of interest to consider a Bayesian setting, to enable inclusion of prior information and assessment of the posterior distribution. Other types of sampling-based methods have been proposed and might be a valuable addition to the current functionality \cite{Donnet2014}. In addition to the limitations discussed above, the dataset structure in the current release of NLMEModeling is rather simple. An extension to a more hierarchical structure (e.g. the Dataset object in Wolfram Mathematica) is currently under consideration as well as automatic import and conversion of NONMEM and Monolix datasets. 

The different examples presented in this paper are meant to serve as an overview of the model-building tools that NLMEModeling provides and the types of models it can process. For additional details regarding function options, methods, and additional examples, we refer the reader to the package documentation in Wolfram Mathematica. One interesting application of SDEMEMs not covered in this paper is to model stochastic behavior in the model parameters, for example the absorption rate in a PK model as presented in the paper by Matzuka \emph{et al.} \cite{Matzuka2016}. In NLMEModeling, this can easily be achieved by introducing a stochastic differential equation to describe the evolution of a specific model parameter of interest.

To conclude, NLMEModeling provides a user-friendly package for performing mixed effects modeling for dynamical systems in Wolfram Mathematica. The development of NLMEModeling is currently active and the package is available on request from \url{http://www.fcc.chalmers.se/software/other-software/nlmemodeling}.
\\\\
\textbf{Acknowledgements}
The authors would like to acknowledge the contribution to the development of the NLMEModeling package that has been done at Fraunhofer-Chalmers Centre for Industrial Mathematics. The authors would especially like to thank Helga Kristín Ólafsdóttir, Simon Berglund Watanabe, and Felix Held.
This work has been partly founded by the Swedish Foundation for Strategic Research by the project Hierarchical Mixed Effects Modeling of Dynamical Systems (Grant no. AM13-0046).
\newpage
\bibliographystyle{naturemag}
\bibliography{mendbib} 
\newpage
\section*{Appendix 1: Simulation Code}
Example 1
\\\\
\framebox[\textwidth]{\includegraphics[]{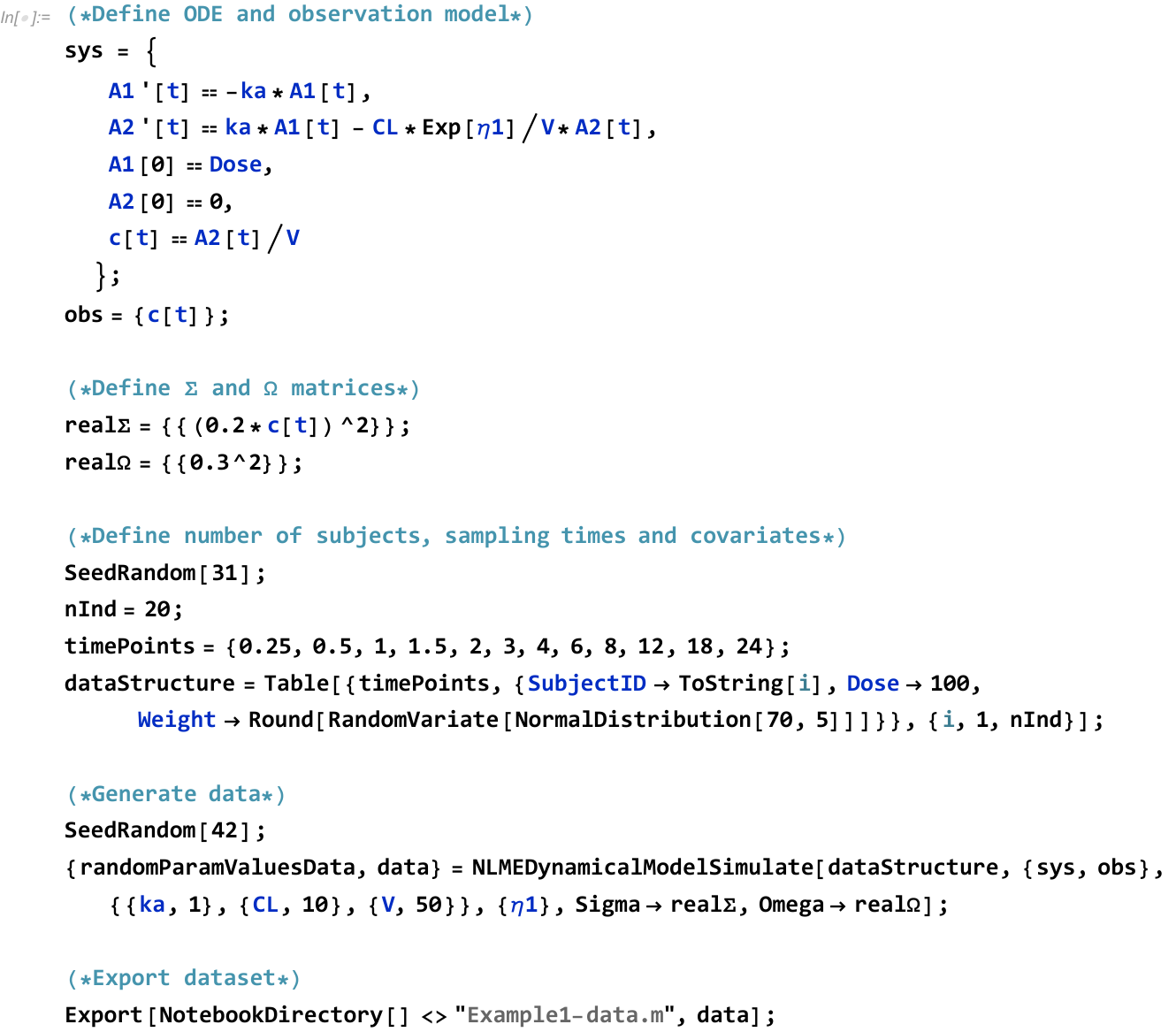}\hfill}
\newpage
\noindent Example 2\\\\
\framebox[\textwidth]{\includegraphics[]{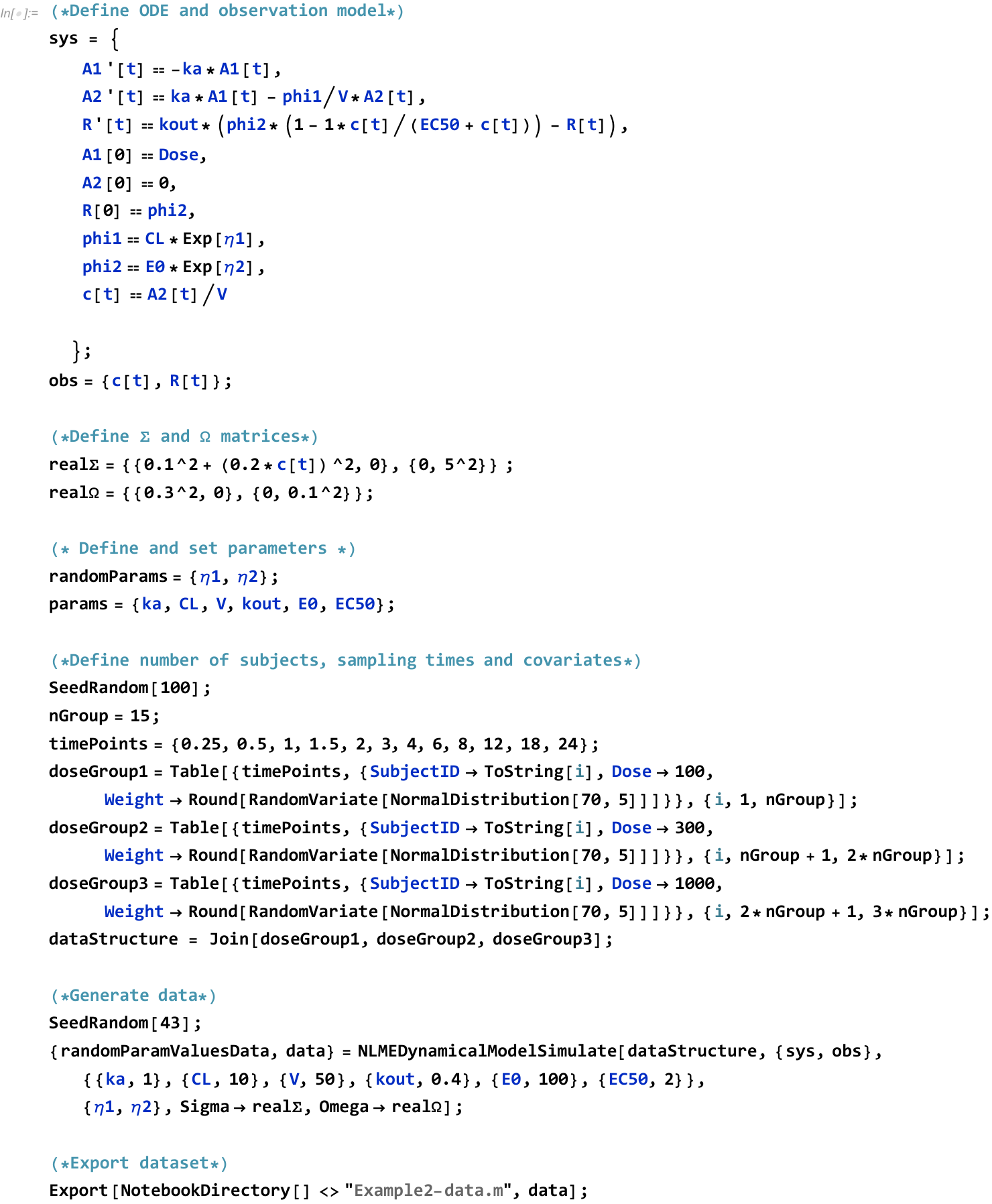}\hfill}
\\\\
\newpage
\noindent Example 3\\\\
\framebox[\textwidth]{\includegraphics[]{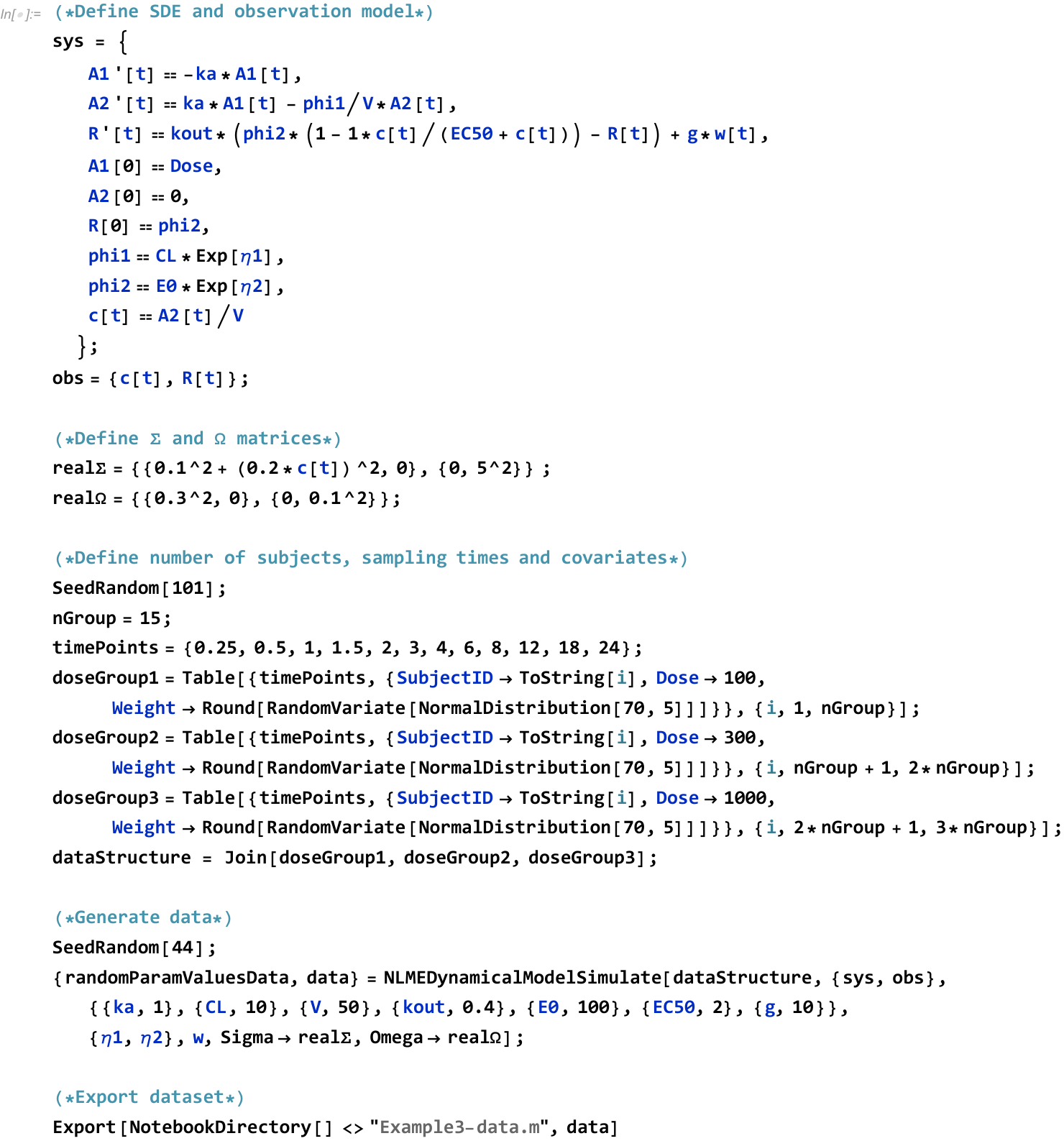}\hfill}
\newpage
\section*{Appendix 2: Derivation of Phi-sensitivites}
The derivation presented here is an extension of the exact gradient method for ordinary differential equations models presented by Almquist et al \cite{Almquist2015}, which is recommended as background material for a full appreciation of the derivation presented in this work.

In the general case, NLME models depends on a vector of fixed effect parameters, $\bm{\theta}$, and a vector of random effect parameters $\bm{\eta}$. As shown in Example 2 \& 3, it might be more convenient to define a set of composite parameters, $\bm{\varphi}$, that can depend on both fixed and random effects, $\bm{\varphi} = \bm{\varphi}(\bm{\theta},\bm{\eta})$. This $\varphi$-parameterization facilitates the definition of for instance normal or log-normal distributed parameters, but also allow for more complex definitions. For maximal flexibility, the NLMEModeling package allows model definitions using $\bm{\theta}$, $\bm{\eta}$, and $\bm{\varphi}$ simultaneously.

Not only does $\varphi$-parameterization allow a more intuitive parameterization of the models, but it also results in a significant simplification of the sensitivity equations that are required for solving the parameter estimation problem. For example, for a normally distributed parameter $\varphi = \theta + \eta$, it is easily seen that the sensitivities of model state variables with respect to $\varphi$, $\theta$, and $\eta$ are identical. Because they are identical, the sensitivity equations for $\theta$ and $\eta$ can be replaced by the sensitivity equations for $\varphi$. In other words, it is possible to compute sensitivities for the original two parameters by only solving the sensitivity equations for the one new parameter. The approach is similar to the MU referencing used by NONMEM for running the FAST version of FOCE \cite{NONMEM74}. Below, we generalize the idea beyond the additive definition of $\varphi$-parameters.

\subsection*{Definitions}
The full vector of fixed effects is defined as the concatenation of the fixed effects used for any $\varphi$-parameter, $\bm{\theta_{\varphi}}$, and their complement, $\bm{\theta_c}$, which only appear on their own,
\begin{equation*}
\bm{\theta} = \bm{\theta_{\varphi}}^\frown\bm{\theta_c}. 
\end{equation*}
Similarly for the random effect parameters, we define
\begin{equation*}
\bm{\eta} = \bm{\eta_{\varphi}}^\frown\bm{\eta_c}. 
\end{equation*}
We also define the parameter vectors
\begin{equation*}
\bm{p_1} = \bm{\varphi}(\bm{\theta_{\varphi}},\bm{\eta_{\varphi}})^\frown\bm{\eta_c} 
\end{equation*}
and
\begin{equation*}
\bm{p_2} = \bm{\theta_c}. 
\end{equation*}
\subsection*{Derivation of relations between sensitivities in the different parameter types}
We let $\bm{x}$ denote the model state variables. The following relations can be derived by applying the chain rule,
\begin{align*}
\frac{d \bm{x}}{d \bm{\eta}} &= \frac{d \bm{x}}{d \bm{p_1}} \frac{d \bm{p_1}}{d \bm{\eta}},\\
\frac{d \bm{x}}{d \bm{\theta}} &= \frac{d \bm{x}}{d \bm{p_1}} \frac{d \bm{p_1}}{d \bm{\theta}} + \frac{d \bm{x}}{d \bm{p_2}} \frac{d \bm{p_2}}{d \bm{\theta}},\\
\frac{d^2 \bm{x}}{d \bm{\eta}^2} &= \left(\frac{d^2 \bm{x}}{d \bm{p_1}^2} \frac{d \bm{p_1}}{d \bm{\eta}}\right)^{T_{132}} \frac{d \bm{p_1}}{d \bm{\eta}}
+ \frac{d \bm{x}}{d \bm{p_1}} \frac{d^2 \bm{p_1}}{d \bm{\eta}^2},\\
\frac{d^2 \bm{x}}{d \bm{\eta} d \bm{\theta}} &= 
\left( \left( \frac{d^2 \bm{x}}{d \bm{p_1}^2} \frac{d \bm{p_1}}{d \bm{\theta}} 
+ \frac{d^2 \bm{x}}{d \bm{p_1} d \bm{p_2}} \frac{d \bm{p_2}}{d \bm{\theta}} \right)^{T_{132}} \frac{d \bm{p_1}}{d \bm{\eta}} \right)^{T_{132}}
+ \frac{d \bm{x}}{d \bm{p_1}} \frac{d^2 \bm{p_1}}{d \bm{\eta} d \bm{\theta}},
\end{align*}
where the tensor and matrix products of the two last equations are performed so that the proper dimensions are matched. This is done using the tensor transpose operator $T_{132}$, which transposes the second and third dimension of a tensor. Note that no "back transpose" is needed in the third equation due to symmetry.

\subsection*{A modified estimation algorithm} 
The existing algorithm can be modified in the following way. The idea is to compute the original first and second order sensitivities of the model state variables, $\bm{x}$, which are required for the likelihood function, but to do this using sensitivities with respect to $\bm{p_1}$ and $\bm{p_2}$. This is achieved by numerically solving the first-order sensitivity equations
\begin{align*}
\frac{d}{dt} \left(\frac{d \bm{x}}{d \bm{p_1}}\right) &= \frac{\partial \bm{f}}{\partial \bm{p_1}}
+ \frac{\partial \bm{f}}{\partial \bm{x}} \left(\frac{d \bm{x}}{d \bm{p_1}}\right)\\
\left(\frac{d \bm{x}}{d \bm{p_1}}\right)(t_0) &= \frac{\partial \bm{x}_0}{\partial \bm{p_1}},\\
\frac{d}{dt} \left(\frac{d \bm{x}}{d \bm{p_2}}\right) &= \frac{\partial \bm{f}}{\partial \bm{p_2}}
+ \frac{\partial \bm{f}}{\partial \bm{x}} \left(\frac{d \bm{x}}{d \bm{p_2}}\right)\\
\left(\frac{d \bm{x}}{d \bm{p_2}}\right)(t_0) &= \frac{\partial \bm{x}_0}{\partial \bm{p_2}},
\end{align*}
and the second order sensitivity equations
\begin{align*}
\frac{d}{dt} \left(\frac{d^2 \bm{x}}{d \bm{p_1}^2}\right) &= 
\frac{\partial^2 \bm{f}}{\partial^2 \bm{p_1}}
+ \left(2\frac{\partial^2 \bm{f}}{\partial\bm{x}\partial\bm{p_1}} + \frac{\partial^2\bm{f}}{\partial\bm{x}^2}\left(\frac{d\bm{x}}{d\bm{p_1}}\right)
\right)^{T_{132}}
\left(\frac{d\bm{x}}{d\bm{p_1}}\right)
+ \frac{\partial \bm{f}}{\partial \bm{x}}\left(\frac{d^2 \bm{x}}{d\bm{p_1}^2}\right)\\
\left(\frac{d^2 \bm{x}}{d \bm{p_1}^2}\right)(t_0) &= \frac{\partial^2 \bm{x}_0}{\partial \bm{p_1}^2},\\
\frac{d}{dt} \left(\frac{d^2 \bm{x}}{d \bm{p_1} d \bm{p_2}}\right) &= 
\frac{\partial^2 \bm{f}}{\partial \bm{p_1} \partial \bm{p_2}}
+\frac{\partial^2 \bm{f}}{\partial \bm{p_1} \partial\bm{x}}\left(\left(\frac{d\bm{x}}{d\bm{p_1}}\right)+\left(\frac{d\bm{x}}{d\bm{p_2}}\right)\right)\\
&+\left(\left(\frac{\partial^2 \bm{f}}{\partial \bm{x}^2}\left(\frac{d\bm{x}}{d\bm{p_2}}\right)\right)^{T_{132}}
\left(\frac{d\bm{x}}{d\bm{p_1}}\right)\right)^{T_{132}}
+ \frac{\partial \bm{f}}{\partial \bm{x}}\left(\frac{d^2 \bm{x}}{d\bm{p_1} d\bm{p_2}}\right)\\
\left(\frac{d^2 \bm{x}}{d\bm{p_1} d\bm{p_2}}\right)(t_0) &= \frac{\partial^2 \bm{x}_0}{\partial\bm{p_1} \partial\bm{p_2}}.
\end{align*}
Note that $d^2\bm{x}/d \bm{p_1}^2$ is a symmetric matrix, which means that either the upper or lower triangular part contains redundant sensitivities. The above sensitivities are subsequently transformed back to the original ones according to the relations in the previous section. The derivatives of $\bm{p_1}$ and $\bm{p_2}$ with respect to the fixed and random effect parameters can be computed symbolically in advance, and the transformation is not expected to be computationally costly. The original sensitivities are then used to compute the population likelihood just like before. 

\subsection*{Impact of $\varphi$-parameterization on the number of sensitivity equations}
The $\varphi$-parameterization introduced above leads to fewer sensitivity equations that have to be solved numerically. The impact in terms of computational time is difficult to predict, mainly since Mathematica's ODE solver to some extent can exploit the fact that many symbolic expressions of the ODE right hand sides are identical, which will be the case for models where a $\varphi$-parameterization is natural. Nevertheless, we can derive expressions for the number of sensitivity equations that are required in the $\varphi$-parameterization and in the original parameterization.

If we let non-bold symbols denote the number of elements in the corresponding bold symbol vectors, the total number of ODEs that have to be solved using the original parameterization is
\begin{equation*}
N^{\rm{old}}_{\rm{in}} = (1+\eta)\,x,
\end{equation*}
for the inner problem, and
\begin{equation*}
N^{\rm{old}}_{\rm{out}} = (1+\eta+\theta+\frac{\eta+\eta^2}{2}+\eta\,\theta)\,x
\end{equation*}
for the outer problem. For the $\varphi$-, or new, parameterization, the numbers are
\begin{equation*}
N^{\rm{new}}_{\rm{in}} = (1+\varphi+\eta_c)\,x,
\end{equation*}
for the inner problem and
\begin{equation*}
N^{\rm{new}}_{\rm{out}} = (1+\varphi+\eta_c+\theta_c+\frac{(\varphi+\eta_c)+(\varphi+\eta_c)^2}{2}+(\varphi+\eta_c)\,\theta_c)\,x
\end{equation*}
for the outer problem.

A favorable situation for $\varphi$-parameterization is when all all random and fixed effects parameters appear in pairs that can be assigned as new $\varphi$-parameters, e.g., $\varphi_1=\theta_1+\eta_1$. In this case $\eta=\theta=\varphi$ and $\eta_c=\theta_c=0$. There are no gains in the inner problem, but the relative number of sensitivity equations in the outer problem is
\begin{align*}
\frac{N^{\rm{new}}_{\rm{out}}}{N^{\rm{old}}_{\rm{out}}} = \frac{(1+\varphi+\frac{\varphi+\varphi^2}{2})\,x}{(1+\varphi+3\frac{\varphi+\varphi^2}{2})\,x},
\end{align*}
which approaches 1/3 in the limit of large $\varphi$.

\subsection*{A strategy for choosing a good parameterization}
The most efficient choice of parameterization in terms of minimizing the number of sensitivity equations is not always trivial. The following three situations are however helpful in order to understand the effects of introducing $\varphi$-parameters.

If a new $\varphi$-parameter can be used to reparameterize the appearance of one random effect and one fixed effect, e.g., $\varphi_1=\theta_1+\eta_1$, then $N^{\rm{new}}_{\rm{in}}$ is unchanged but $N^{\rm{new}}_{\rm{out}}$ is decreased by $(1+\phi+\eta_c)\,x$. This kind of reparameterization is therefore desirable as the number of sensitivity equations always decrease. If a new $\varphi$-parameter is used to reparameterize a fixed effect only, then $N^{\rm{new}}_{\rm{in}}$ increases by $x$, and $N^{\rm{new}}_{\rm{out}}$ increases by $\theta_c\,x$. If, on the other hand, a new $\varphi$-parameter is used to reparameterize a random effect only, the number of equations remain constant for both the inner and outer problem.

There may be other more complex situations that require careful consideration as to which parameterization that is better. As a general recommendation, one or a combination of several $\varphi$-parameters should only be used to replace a larger number of fixed and random effects.

\subsection*{Extension to additional parameter types}
The idea presented above could also be extended to include inter-occasion random effects, $\bm{\kappa}$. In this scenario, $\bm{\varphi}$ may also depend on $\bm{\kappa}$, $\bm{\varphi} = \bm{\varphi}(\bm{\theta},\bm{\eta},\bm{\kappa})$. With $\bm{\kappa} = \bm{\kappa_{\varphi}}^\frown\bm{\kappa_c}$, we redefine $\bm{p_1} = \bm{\varphi}(\bm{\theta_{\varphi}},\bm{\eta_{\varphi}},\bm{\kappa_{\varphi}})^\frown\bm{\eta_c}^\frown\bm{\kappa_c}$. The sensitivities of the model state variables with respect to the inter-occasion random effects can now be obtained in a similar way as outlined above for the standard random effects, e.g.,
\begin{equation*}
\frac{d \bm{x}}{d \bm{\kappa}} = \frac{d \bm{x}}{d \bm{p_1}} \frac{d \bm{p_1}}{d \bm{\kappa}}.
\end{equation*}
If $\bm{\kappa_c}$ is empty, this requires no new sensitivity equations at all.

\subsection*{Removing trivial sensitivity equations}
The estimation algorithm can be improved by removing trivial sensitivity equations whose solution is time independent, i.e., constant. In the $\varphi$-parameterization framework, this works in the following way. 

First, a concatenated vector of the original state variables and the sensitivity state variables is formed for the inner problem,
\begin{equation*}
\bm{s_{1}} = \bm{x}^\frown\vect\left(\frac{d \bm{x}}{d \bm{p_1}}\right),
\end{equation*}
where $\vect()$ denotes the flattening of a matrix to a vector. Similarly, we have for the outer problem
\begin{equation*}
\bm{s_{2}} = \bm{x}^\frown\vect\left(\frac{d \bm{x}}{d \bm{p_1}}\right)^\frown\vect\left(\frac{d \bm{x}}{d \bm{p_2}}\right)^\frown\vect\left(\frac{d^2 \bm{x}}{d \bm{p_1}^2}\right)^\frown\vect\left(\frac{d^2 \bm{x}}{d \bm{p_1} d \bm{p_2}}\right).
\end{equation*}
The complete systems of ODEs for the inner and outer problem can be then be compactly written
\begin{align*}
\frac{d\bm{s_1}}{dt}  &= \bm{\mathrm{g_1}}(\bm{s_1},t,\bm{p_1})\\
\bm{s_1}(t_0) &= \bm{s}_{\textbf{1}0},
\end{align*}
and
\begin{align*}
\frac{d\bm{s_2}}{dt}  &= \bm{\mathrm{g_2}}(\bm{s_2},t,\bm{p_1},\bm{p_2})\\
\bm{s_2}(t_0) &= \bm{s}_{\textbf{2}0}.
\end{align*}

Second, for both the inner and the outer problem, a so called connectivity matrix, $\bm{\mathrm{M}}$, is defined. For the inner problem, an element $\bm{\mathrm{M}}_{ij}$ of the connectivity matrix is equal to 1 if the $j$th element of $\bm{s_1}$ appears in the $i$th element of $\bm{\mathrm{g_1}}$, otherwise it is 0. In addition, all elements on the diagonal of $\bm{\mathrm{M}}$ are set to 1. The connectivity matrix is defined in the corresponding way for the outer problem. The connectivity matrix is then iteratively applied to a vector $\bm{z}$
\begin{equation*}
\bm{z}_{n+1}=\sgn(\bm{\mathrm{M}}\,\bm{z}_{n}),
\end{equation*}
until $\bm{z}_{n+1}=\bm{z}_{n}$. The vector $\bm{z}$ is initialized by setting the $i$th element to 0 if the $i$th element of $\bm{\mathrm{g_1}}$ (after replacing in variables with their initial conditions) is identically equal to 0 at $t_0$, or setting it to 1 otherwise. The point of the above idea is to identify which (sensitivity) variables that initially are changing, and then see how these changes propagate through the inter-connected system of ODEs (if a variable initially is changing, it may continue to change without influence of other state variables, and therefore we must set $\bm{\mathrm{M}}_{ii}=1$). Any zeros in the fixed point iterate of $\bm{z}$ shows that these variables have a trivial constant solution.

Third, when the trivial equations have been identified, the appearance of the corresponding variables in the other equations must be replaced by their constant solutions.

\end{document}